\documentclass[manuscript]{acmart}

\usepackage{tabu}
\usepackage{booktabs}
\usepackage{mathrsfs}
\usepackage[normalem]{ulem}
\usepackage{xspace}
\usepackage{amsmath}
\usepackage{enumitem}
\usepackage{algorithm}
\usepackage{algorithmicx}
\usepackage[noend]{algpseudocode}
\usepackage{xspace}
\usepackage{graphicx}
\usepackage{duckuments}
\usepackage{multirow}
\usepackage{xcolor}
\usepackage{makecell}
\usepackage{array}
\usepackage{longtable}

\definecolor{newred}{HTML}{EB5757}
\definecolor{newblue}{HTML}{2D9CDB}
\definecolor{newgreen}{HTML}{27AE60}

\newcommand{\eg}{\textit{e.g.}}

\newcommand{\ie}{\textit{i.e.}}
\newcommand{\etal}{\textit{et al.}}
\newcommand{\tool}{\emph{VirtuWander}\xspace}

\newcommand{\stage}{\textcolor[HTML]{EB5757}{\textbf{Stage}}\xspace}
\newcommand{\stages}{\textcolor[HTML]{EB5757}{\textbf{Stages}}\xspace}
\newcommand{\task}{\textcolor[HTML]{2D9CDB}{\textbf{Task}}\xspace}
\newcommand{\tasks}{\textcolor[HTML]{2D9CDB}{\textbf{Tasks}}\xspace}
\newcommand{\info}{\textcolor[HTML]{27AE60}{\textbf{Information}}\xspace}
\newcommand{\response}{\textcolor[HTML]{F2994A}{\textbf{Feedback}}\xspace}
\newcommand{\q}[1]{\textit{``#1''}}

\newcommand{\voice}{\textit{voice}\xspace}
\newcommand{\avatar}{\textit{avatar}\xspace}
\newcommand{\abs}{\textit{text window}\xspace}
\newcommand{\map}{\textit{minimap}\xspace}
\newcommand{\mirror}{\textit{virtual screen}\xspace}
\newcommand{\highlight}{\textit{highlight}\xspace}
\newcommand{\sign}{\textit{signpost}\xspace}

\newlist{customenum}{enumerate}{1}
\setlist[customenum,1]{label=\textbf{C\arabic*}}

\usepackage[normalem]{ulem}

\AtBeginDocument{%
  \providecommand\BibTeX{{%
    \normalfont B\kern-0.5em{\scshape i\kern-0.25em b}\kern-0.8em\TeX}}}

\setcopyright{acmcopyright}
\copyrightyear{2018}
\acmYear{2018}
\acmDOI{XXXXXXX.XXXXXXX}

\acmConference[Conference acronym 'XX]{Make sure to enter the correct
  conference title from your rights confirmation emai}{June 03--05,
  2018}{Woodstock, NY}
\acmPrice{15.00}
\acmISBN{978-1-4503-XXXX-X/18/06}

\acmSubmissionID{7938}

\begin{document}

\title[\tool]{\tool: Enhancing Multi-modal Interaction for Virtual Tour Guidance through Large Language Models}

\author{Zhan Wang}
\email{zwang834@connect.hkust-gz.edu.cn}
\affiliation{%
  \institution{The Hong Kong University of Science and Technology (Guangzhou)}
  \country{China}
}

\author{Linping Yuan}
\affiliation{%
  \institution{The Hong Kong University of Science and Technology}
  \country{Hong Kong SAR, China}
}

\author{Liangwei Wang}
\affiliation{%
  \institution{The Hong Kong University of Science and Technology (Guangzhou)}
  \country{China}
}

\author{Bingchuan Jiang}
\email{jbc021@163.com}
\affiliation{%
  \institution{Strategic Support Force Information Engineering University}
  \country{China}
}

\author{Wei Zeng}
\affiliation{%
  \institution{The Hong Kong University of Science and Technology (Guangzhou)}
  \country{China}
}


\begin{abstract}

Tour guidance in virtual museums encourages multi-modal interactions to boost user experiences, concerning engagement, immersion, and spatial awareness.
Nevertheless, achieving the goal is challenging due to the complexity of comprehending diverse user needs and accommodating personalized user preferences. 
Informed by a formative study that characterizes guidance-seeking contexts, we establish a multi-modal interaction design framework for virtual tour guidance.
We then design \tool, a two-stage innovative system using domain-oriented large language models to transform user inquiries into diverse guidance-seeking contexts and facilitate multi-modal interactions.
The feasibility and versatility of \tool are demonstrated with virtual guiding examples that encompass various touring scenarios and cater to personalized preferences.
We further evaluate \tool through a user study within an immersive simulated museum. 
The results suggest that our system enhances engaging virtual tour experiences through personalized communication and knowledgeable assistance, indicating its potential for expanding into real-world scenarios.

\end{abstract}




\maketitle

\section{Introduction}
\label{sec:introduction}
A virtual museum refers to a digital entity encompassing a physical museum's characteristics~\cite{schweibenz2019virtual}, which can be visited on various devices like augmented reality (AR) and virtual reality (VR) glasses~\cite{dasgupta2021redefining,woolley2020virtual, li2021cubemuseum}.
With the rapid advancement of AR \& VR technology, virtual museums are currently experiencing increasing interest in various domains such as education~\cite{tserklevych2021virtual, daniela2020virtual} and cultural heritage~\cite{pirbazari2022recordim, esmaeili2018conceptual}.
These applications leverage diverse interaction methods in virtual environments, such as voice narration, text guidance, and avatar interaction, to emulate human tour guides for completing tasks like knowledge explanation~\cite{richards2012agent,aiello2019virtual} and spatial navigation~\cite{ferracani2017natural,tsitseklis2023recbot}.
While finding a good tour guide service in the real world can be challenging, VR addresses this issue by offering diverse, flexible, and virtual assistance~\cite{tatli2023analysis} and thus boosts the audience visiting experience.
To make a touring experience more personalized, interactive, and knowledgeable, virtual tour guidance is mainly considered and designed from the aspects of immersion~\cite{zidianakis2021invisible, esmaeili2018conceptual,zou2021value}, engagement~\cite{tsita2021approach, zou2021value}, and spatial awareness~\cite{sylaiou2010exploring,chiu2000museum}.
However, this demands designing tour guidance with natural interaction, responsive information delivery, and satisfying personalized needs.

Despite its importance, the current research on designing tour guidance in virtual museums remains limited.
Existing studies primarily focus on non-flexible or semi-flexible interactions for museum tour guidance~\cite{tatli2023analysis}, such as directing visitors along predefined routes~\cite{tsitseklis2023recbot} or presenting pre-written commentary~\cite{rzayev2019fostering}. 
Consequently, these approaches yield constrained interactions, offering only a restricted set of functionalities and templated responses.
In contrast, achieving effective tour guidance demands the capability to accommodate general guidance-seeking scenarios and the flexibility to customize guides according to individual preferences.
Nonetheless, this remains a nontrivial task, significantly hindered by the expansive exploration space encompassing diverse environmental factors within a virtual museum, personalized guidance needs for interacting with the environment, and the integration of multi-modal interaction methods.
In particular, a virtual museum typically simulates an environment where artworks are positioned in various spatial positions. The artworks contain diverse information, including details about their authors and content, presented from multiple perspectives.
Users may desire a general tour to view all artworks or prefer to explore specific artworks in greater detail. 
These distinct types of information can be conveyed to users through various guidance feedback.
Simply transferring guidance interactions from the physical world, such as using voice or text, is insufficient for users to fully engage with virtual environments~\cite{zhu2023virtual}.

Large language models (LLMs), such as GPTs~\cite{openai2023gpt4, openai2022chatgpt} and LLaMAs~\cite{touvron2023llama,touvron2023llama2}, have demonstrated their ability to understand user intent and engage in long contextual multi-turn dialogues.
They are widely applied in various downstream tasks~\cite{park2023generative,wang2023enabling,roziere2023code}, where system builders leverage prompts with a limited number of examples relevant to the target tasks to control LLMs.
This generalization holds the promise of meeting the personalized guidance needs of the general audience in virtual museums.
However, it still remains unclear about the feasibility of leveraging LLMs seamlessly for tour guidance in virtual museums.
Limited research has been conducted to explore how LLMs can be adapted to intuitively and individually support tour guidance tasks in this context.
First, LLMs' output modality is confined exclusively to natural language, which may not be the most intuitive and effective means for gathering information and establishing user-environment relationships in the physical world. 
Furthermore, to generate user-centric responses tailored for virtual museums, it is necessary to augment LLMs' knowledge with implicit environmental information and specific user requirements.
Therefore, this paper primarily investigates the viability of multi-modal feedback mechanisms for customizing LLMs to enhance the touring experience in virtual museums.

This paper presents the design of \tool, an innovative system that leverages LLMs to enhance multi-modal interactions for virtual tour guidance.
We take an initial application in art museums due to their typically solitary user experience and the restricted accessibility to precious artworks~\cite{hornecker2019human}.
We first conduct a formative study to interview general users' guidance-seeking contexts within various touring scenarios and categorize contexts based on when users require guidance (\stage), what implicit environmental information is necessary to provide guidance (\info), and the specific guidance users require for their tasks (\task).
Drawing from commonly encountered guided tour scenarios, we establish a comprehensive framework comprising seven primary multi-modal guidance (\response) that users generally anticipate LLMs to facilitate.
Following this design framework, we develop an innovative \tool system to enable five multi-modal guidance feedback combination designs with natural language inputs in virtual museums.
To accommodate personalized user needs in the context of the virtual museum, we leverage a pack-of-bots strategy, with each LLM-based chatbot embellished with domain-specific knowledge of task descriptions, context-based constraints, few-shot examples, contextual information, and human-environment connection.
We demonstrate the feasibility and versatility of \tool through three virtual guiding examples including a thematic tour exploration, a single artwork exploration, and a personal tour customization, encompassing diverse tour contexts and addressing personalized user requirements.
We also conduct a user study in a simulated virtual museum to gain user feedback for each LLM-empowered multi-modal guidance design and the system \tool as a whole.
The evaluation results validate the effectiveness of our approach in facilitating personalized communication, delivering knowledgeable assistance, and enhancing engaging virtual tour experiences, which indicates its future potential for real-world tour guidance.

In summary, our main contributions consist of the following points: 
\begin{itemize}
    \item 
    We summarize a design framework from a formative study for LLM-empowered multi-modal feedback to enhance various tour contexts with interactive guidance.
    \item We introduce \tool, a voice-controlled prototype that demonstrates five interaction designs within a simulated virtual museum following our design framework. 
    \tool incorporates a two-stage strategy, involving context identification and feedback generation, to leverage LLM to bridge users' natural language input and multi-modal feedback.
    \item We evaluate the capabilities, potential, and limitations of LLM-enhanced multi-modal interactions for guided tour experiences through multiple showcases and a user study.
\end{itemize}

\section{Background and Related Work}
\label{sec:related-work}

This research draws on prior work on understanding the requirements of multi-modal interactions in virtual museums, challenges for designing multi-modal interactions for tour guidance, and progress in LLM-based interaction design.

\subsection{Virtual Museums}

Museums are identified as a place of "free-choice learning"~\cite{falk2000learning,falk2006living}, allowing visitors to tailor various experiences to their changing motivations and interests.
Such visitor experience of museums is deeply influenced by three dimensions: physical context (\eg, exhibited objects), social context (\eg, interactions with other people), and personal context (\eg, preference and knowledge)~\cite{falk1992museum}.
However, physical museums often face constraints related to space, time, and access to valuable collections~\cite{waern2022hybrid}.
In contrast, virtual museums are increasingly utilizing digital technologies to enhance the audience reach and complement physical exhibitions~\cite{hornecker2019human}.
A virtual museum represents a digital entity that serves as a multimedia tool for disseminating museum information~\cite{styliani2009virtual,pujol2012virtual}.
Due to their digital medium and cultural characteristics, virtual museums gain wide applications in various domains like education~\cite{tserklevych2021virtual, daniela2020virtual} and cultural heritage preservation~\cite{aiello2019virtual, esmaeili2018conceptual}. 
These applications demonstrate that virtual museums can be deployed through various digital devices and technologies, including the Web~\cite{dasgupta2021redefining}, VR~\cite{ferracani2017natural, kersten2017development}, and AR~\cite{woolley2020virtual, li2021cubemuseum}.
The overarching goal of virtual museums is consistently to boost the user tourism experience through personalization, interaction, and knowledge enrichment~\cite{schweibenz2019virtual}.
Specifically, the advances in VR technologies offer great opportunities to create more accessible virtual museums for artworks and innovative presentation formats~\cite{sooai2017virtual}.
For instance, Tsita \etal~\cite{tsita2021approach} showed that virtual museums can encourage user exploratory behaviors in a virtual contemporary art exhibition and enrich a broad audience's knowledge of artwork, by employing more painting-stimulated interactions including static displays and 3D animations.
Recent studies have also been devoted to enhancing the user experience of museum tourism in VR from various aspects, including engagement~\cite{tsita2021approach,sylaiou2010exploring}, immersion~\cite{sylaiou2020avatars, zidianakis2021invisible,sooai2017virtual}, personalization~\cite{dattolo2008visualizing, bonis2009platform}, spatial awareness~\cite{tsitseklis2023recbot}.

However, when compared to traditional web-based devices, designing a user-friendly visitor experience in VR poses greater challenges due to the complexity of interactions~\cite{tsita2021approach} and disconnection from the physical world~\cite{chiu2000museum}.
Existing studies have primarily focused on specific usage scenarios characterized by limited interactions, making it challenging to provide comprehensive guidance in virtual museums.
For instance, in the case of the Virtual Artifact application~\cite{sooai2017virtual}, virtual artifacts are placed within a virtual environment, and a predetermined tour route is defined for visiting all the artifacts.
However, such restricted tour guidance interactions are insufficient for users to fully immerse themselves in the virtual museum experience~\cite{zhu2023virtual}.
To fill the gap, this work is dedicated to designing a more flexible and universally applicable guidance interaction approach to enhance the user tour experience in virtual museums.
We aim to support diverse visitor motivations, from curiosity-driven exploration to seeking specialized knowledge or unique experiences, offering a flexible and responsive virtual visiting experience that adapts to individual needs and interests.
In particular, we take an initial step from art museums due to the isolated user experience in art museums and limited accessibility to precious artworks~\cite{hornecker2019human}.
Our contribution entails a comprehension of the design space for multi-modal interactions in virtual tour guidance, and the introduction of an LLM-based approach to accomplish the goal.

\subsection{Multi-modal Interaction for Tour Guidance}

Multi-modal interaction integrates various input modalities and diverse sensory feedback mechanisms to enable an immersive and intuitive user experience. 
In virtual tour guidance, incorporating a wide range of input modalities, such as voice~\cite{alghofaili2019lost, von2020podoportation} and body motion~\cite{hombeck2023tell, li2019research, rantamaa2022evaluation}, leads to more natural and engaging interactions that mirror human communication and multi-sensory experiences in the real world~\cite{hornecker2019human}.
Likewise, multi-modal feedback can encompass auditory cues~\cite{Monastero2016wandertroper}, visual feedback, behavioral responses from haptic feedback ~\cite{jiang2021douleur,levent2014multisensory}, smell and taste sense~\cite{Obrist2017master,claisse2018multisensory}, and various combinations thereof~\cite{baxter2021you, zhu2023virtual,vi2017midair}. 
These diverse input modalities and output feedback offer tremendous flexibility for creating immersive VR experiences.
Such multi-modal interactions are also identified as helpful to resonate with the aesthetics~\cite{davis2015don} and convey evocative narratives around artworks~\cite{hornecker2019human}.
However, the rich variety of multi-modal interactions also increases complexities and difficulties when promoting visitor experiences in virtual museums.

Considering the typical tour guidance experience in the real world, where guides and tourists primarily communicate through spoken language, we have opted to exclusively utilize voice as the input method for this work.
Voice communication is a human instinct, and its effectiveness in enhancing emotional impact and participation has also been confirmed in VR voice-based interactions~\cite{osking2019enhancing, ferracani2017natural}.
Due to its hands-free, intuitive, and efficient features, voice-based control has been gaining popularity in locomotion tasks~\cite{baxter2021you, hombeck2023tell} or as a means of virtual avatars and agents as guides.
Having established this decision, a primary challenge for this work lies in selecting suitable output feedback mechanisms for tour guidance in virtual museums.
Although multi-modal interaction mechanisms have been largely experimented with in physical museums~\cite{levent2014multisensory, vi2017midair, claisse2018multisensory}, there is limited knowledge about multi-modal interaction design in virtual museums.
It is inappropriate to directly shift those interaction mechanisms into the virtual environments due to the great immersion and flexibility for interaction in virtual reality~\cite{tost2007exploring} and higher user expectations for interesting and novel interaction designs~\cite{waern2022hybrid}.
To fill this gap, we initiated a preliminary study aimed at understanding user preferences for multi-modal feedback across various guiding contexts in virtual museums. 
This study revealed preferences for a consistently present virtual avatar and voice narration, as well as the occasional inclusion of visual cues and spatial movements. 
The design space serves as the foundation for building the LLM-based multi-modal interaction design.

\subsection{LLM-based Interaction Design}

The profound generative and interpretive potential of large language models (LLMs) marks a transformative approach to interaction design.
LLMs have demonstrated their capabilities in understanding user intent and supporting extensive dialogues with a wealth of commonsense world knowledge~\cite{openai2023gpt4, openai2022chatgpt,touvron2023llama,touvron2023llama2}.
Studies have presented a new norm to leverage LLMs for different downstream tasks, such as creating generative agents~\cite{park2023generative}, assisting programming~\cite{roziere2023code}, and improving communication~\cite{wang2023enabling}, all through a few-shot learning approach known as prompting engineering.
Few-shot learning involves providing a pre-trained model with a small number of task-specific examples and allowing it to generalize across various tasks without altering the model parameters~\cite{liu2023pre}.
Such generalization opens up possibilities for extending LLMs into VR.
For example, Roberts \etal~\cite{roberts2022steps} proposed prompted-based methods for code generation to support AI-assisted co-creation of 3D objects in VR.
Project Mellon by NVIDIA integrates speech AI and LLMs to simplify interactions in immersive environments~\cite{NVIDIA2023ProjectMellon}.

Recent studies such as VELMA~\cite{schumann2023velma} and LM-Nav~\cite{shah2023lm} have demonstrated the potential of harnessing LLMs for real-world navigation or spatial awareness related tasks.
These works utilize LLMs to compute navigation paths in static environments offline.
Nevertheless, no research has explored the specific context of virtual museums and considered how to leverage LLMs to enhance visitor experiences in virtual environments.
Despite the promise of LLMs, it remains challenging to achieve effective interaction designs within virtual museums that can adapt to versatile user intentions and offer in-situ feedback in real-time.
The tasks are more intricate, involving interactions with dynamic user inputs that need to be processed responsively in VR environments.
To fulfill this objective, we have developed a two-stage approach with a pack-of-bots strategy to efficiently address complex user requirements within a limited time frame.
\section{Design Space}
\label{sec:design-space}

To inform the design of our system, we conducted a formative study (Sect.~\ref{ssec:formativestudy}) to understand user needs for virtual tour guidance.
We selected three virtual art museums as the study environment for 12 experienced virtual museum visitors.
The findings show that visitors seek guidance in different contexts, which can be described by stage, information, and task.
On this basis, we propose a design framework that characterizes visitors' guidance-seeking contexts and expected multi-modal guidance feedback (Sect.~\ref{ssec:designframework}).
Then we identified the five most expected feedback combinations (Sect.~\ref{ssec:context_walkthrough}) and implemented them in our system (Sect.~\ref{sec:prototype}).

\subsection{Formative Study}
\label{ssec:formativestudy}
\subsubsection{Setup}

\textbf{Participants.}
To recruit participants, we shared our study description within online VR player communities and selected participants who had previously visited VR museums and owned VR headsets.
Our study included 12 participants (P1-P12; Age: 18-32; 5 females).
They reported an average familiarity with VR of 4.5 (SD=2.3, MAX=7, MIN=1) on a 7-point Likert scale.
On average, they had visited 3 (SD=2.7, MAX=10, MIN=1) virtual museums.

\textbf{Stimuli.} 
We selected and purchased three virtual museums with varying spatial layouts and exhibit types (\eg, 2D paintings and 3D sculptures) from an online 3D model store.
One museum served for testing purposes and the other two provided specific scenarios for us to observe the participants' virtual tour activities and understand their needs.
We wrapped these three museums into three VR applications, providing basic functions such as continuous movement and teleportation.
Prior to each study, participants installed the VR applications and ensured that they could find a safe sitting place, cast their VR views to computer screens, and share their screens via Zoom.

\textbf{Procedure.}
We conducted the study individually with each participant via Zoom, recording the sessions after obtaining their consent.
After briefly introducing our study goals and protocol, we invited them to have tours in the virtual museums and think aloud about their actions and any guidance they sought.
Specifically, we first conducted a short tutorial session with the testing museum, instructing the participants on teleportation using buttons and movement with a joystick.
We also encouraged them to practice thinking aloud their thoughts. After the training session, they toured the second museum, after which we conducted a brief interview based on their think-aloud thoughts. The visit and short interview were repeated with the third museum. Throughout their exploration, participants had the freedom to navigate the virtual museums and engage with the artworks. We could observe their activities through the cast view on our computer screens. During each interview, they reflected on their tours, discussed their guidance needs, and imagined interactions with an intelligent agent powered by an LLM model to enrich their tour experience.

\textbf{Data Analysis.}
After transcribing the think-aloud thoughts and interviews, we performed a thematic analysis~\cite{braun2019reflecting}. Two authors first performed independent coding by reading transcriptions, marking the sentences related to the needs for guidance and expected interactions with an intelligent LLM-based agent, assigning codes for each sentence, and grouping similar codes with affinity diagramming. After that, the two authors compared and discussed the codes to achieve a single coding scheme.

\subsubsection{Findings}

Our findings show that the participants sought guidance in many different contexts, which we characterize by \stage, \info, and \task. The results also indicate a preference for voice-only control alongside multi-modal guidance \response.

\textbf{Visitors seek guidance in three~\stages.}
First, at the \textbf{beginning} of their visit, all visitors (N=12) felt the need for an initial orientation. They expected guidance similar to \q{a map in a shopping mall that overviews the entire space} (P1). This desire arose from \q{unfamiliarity with the environment} (P3) and a wish to \q{become acquainted with the layout and pinpoint potential exhibits of interest} (P9).
As their visits were \textbf{in progress}, they (N=12) frequently chose artworks to engage with based on their immediate surroundings and initial impressions. Nonetheless, they expressed a desire for added guidance to make more informed choices, echoing P12's sentiment, \q{while immediate surroundings allow me to quickly connect with exhibits that caught my eye, there was a risk of missing out on other artworks that might not have been immediately visible}. This highlighted a desire for \q{a seamless transition between exhibits without missing any} (P2) and an efficient tour guide to \q{cover all exhibits of interest} (P5).
When their visit was \textbf{ending}, most participants (N=8) expressed a need for reflection and review. They wished for a \q{history or visual documentation of their journey} (P4), allowing them to \q{reminisce and ensure they had a comprehensive experience} (P7). This could be in the form of \q{photos they had taken or an automated log} (P10).

\textbf{Visitors need three types of~\info.}
Based on the responses, the information desired by the participants can be categorized into three groups.
First, all visitors (N=12) required \textbf{spatial} information, which was related to the physical layout and orientation of the exhibits within the museum. To obtain spatial information, visitors might \q{roam around to get a sense of the layout} (P1). Such information also aided them in choosing the next exhibit to view based on \q{the distance between the exhibits and me} (P8).
Second, all visitors (N=12) wanted \textbf{semantic} information that pertained to the actual content and details of the exhibits, such as the attributes and stories behind the exhibits. Semantic information was important for visitors' understanding. For example, P10 explained how the name and introduction of an artwork guide their viewing. Visitors also made viewing choices based on different semantic information but with different preferences. For example, P9 expressed that, \q{the visual style and colors were the main factors attracting me to consume an exhibit}. 
Lastly, most visitors (N=9) also expressed the need for \textbf{social} information, which was related to other visitors' interactions with and impressions of the museum.
The participants expressed their curiosity about other visitors' activities and emotional reactions because \q{shared experiences and collective discoveries could enhance the overall museum journey} (P10). Specifically, P6 and P7 both expressed interest in \q{what others find intriguing about an exhibit}. P1 and P4 highlighted the utility of seeing others' notes, photos, or interactions with an exhibit, which could serve as a guide. Compared to activities related to individual persons, the participants were more interested in communal engagement and used them differently. For example, P9 and P12 might choose a route most other visitors took while P5 wanted to \q{view the least seen exhibits}.

\textbf{Visitors want to enhance their experiences by performing three~\tasks.}
When the participants sought the above three types of information, they had different purposes and performed three different tasks with the information.
The first was \textbf{information enhancement} mentioned by all visitors (N=12), referring to visitors wanting to enrich their understanding of the exhibits they were engaging with. As mentioned above, they wanted semantic information such as \q{the artwork's name and introduction} (P10) to help them derive meaning, social information such as \q{how others rated the exhibit} (P5), and \q{how others observed a 3D sculpture from different perspectives} (P4) to find something interesting, and spatial information to have an overall understanding of the museums. 
Specifying \textbf{personalized preference} was the second task where some visitors (N=7) aimed to tailor their experience based on their interests. Many participants showed personal habits and preferences that determined their exploration path. Some made decisions based on set habits, such as \q{always opting for a counter-clockwise viewing pattern} (P2).  Some visitors optimized their path to \q{avoid doubling back} (P12) or \q{ensuring they didn't miss out on viewing any piece} (P7). Another shared the importance of \q{noting which artworks I had already seen and which ones remained} (P11).
\textbf{Navigation} was another key task for all visitors (N=12). 
When faced with multiple paths or directions, visitors required clear guidance, especially if they were unsure of their next move. P4 expressed \q{the need for direct navigation, especially when confronted with a multi-directional choice}. P3 recognized the importance of \q{familiarizing myself with the environment for better route planning based on their spatial cognition and the currently visible artworks.} Additionally, locating oneself and understanding one's trajectory, especially when wanting to \q{return to a starting point} (P1), became crucial in large museums.

\textbf{Visitors expected seven modes of~\response.}
When envisioning interactions with an LLM-powered intelligent agent to enhance the tour experience, all participants (N=12) expressed a preference for voice control due to its naturalness. 
Each participant mentioned some expected feedback they hoped to receive after issuing voice commands and we summarized a total of seven different modes of feedback that were most frequently suggested across various contexts.
The first feedback was \textbf{avatars} (N=12). They expressed the need for visual cues, especially avatars that could guide them to consume the exhibits. Some participants mentioned some interesting usages of avatars. For example, P5 said, \q{Maybe a visitor discovered special perspectives, such as viewing through a hollow cube to observe the large 3D installation behind it. How about showing their unique postures with avatars to inform others  who may not notice?}
Second, \textbf{voice} assistance (N=12), akin to a real museum tour guide, was desired, such as \q{the avatars could audibly introduce artworks} (P6).
The third need was \textbf{text window} (N=9). Some participants desired supplemental textual details to reinforce the voice or visual elements. For example, P1 and P9 hoped for detailed textual information on artworks besides long voices. P2 took it further, \q{I wish for a concise summary of my viewing experience generated by LLM because self-compilation is tedious.} 
\textbf{Minimaps} (N=10) was the fourth mode, which can act as an official navigation guide (P8) or a tool to \q{tracing my museum journey to ensure no exhibit was overlooked} (P7).
Fifth, clear \textbf{signposts} (N=8) which could work together with minimaps were wanted. P11 imagined directional arrow indicators to avoid getting lost.
Sixth, \textbf{highlight} (N=4) features on artworks were a favorite. For example, P10 said, \q{I hope for specific highlights on artworks to emphasize areas of importance or significance, potentially reflecting the curator's intent or focus areas that might otherwise be overlooked by a casual observer like me.}
Lastly, the concept of a \textbf{virtual screen} (N=2) for intuitive artwork comparisons was brought up. When P4 saw three pictures with the same themes, she wanted \q{a tool to aid me in drawing direct connections, understanding similarities or contrasts, and achieving a deeper appreciation of the artworks.}
The participants did not express their preferences for the seven feedback modes but suggested that each feedback mode should appear at an appropriate time, based on the stage they visited, the type of information they requested, and the task they wanted to perform. This drives us to develop a design framework (Sect.~\ref{ssec:designframework}).

\begin{figure}[t]
    \centering
    \includegraphics[width=0.99\linewidth]{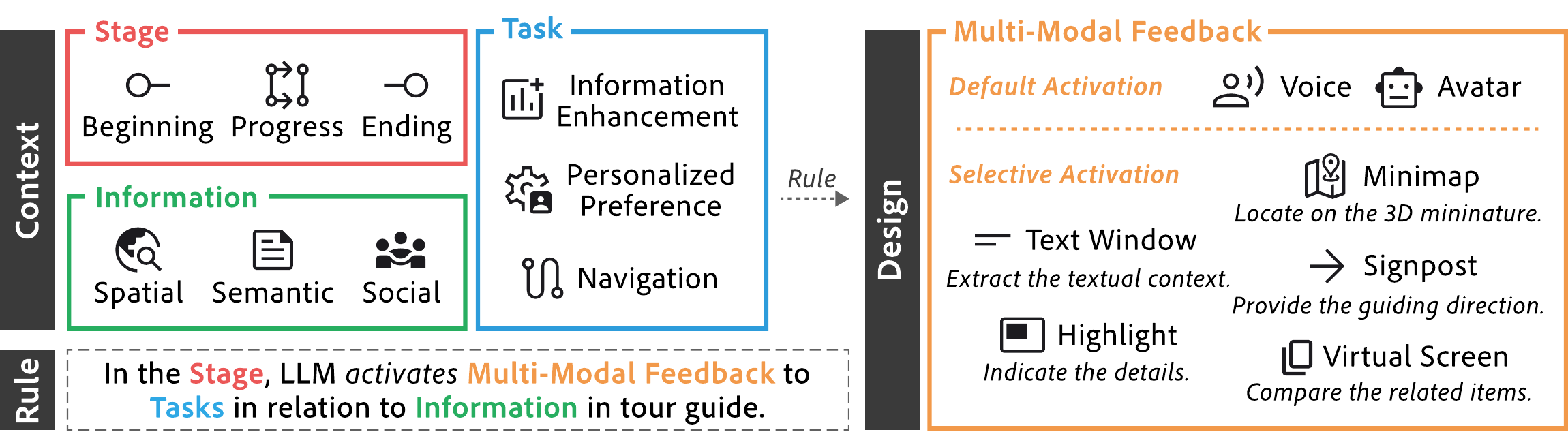}
    \caption{Design framework for LLM-based multi-modal feedback within various guidance-seeking contexts in virtual tour experiences.}
    \vspace{-3mm}
    \label{fig:design}
\end{figure}

\subsection{Design Framework}
\label{ssec:designframework}
To better formulate our findings and inform the system design, we propose a design framework to build the relationships among the guidance-seeking contexts and expected multi-modal feedback -- 
\q{In the~\stage, LLM activates~\response to~\tasks in relation to~\info in tour guidance}. 
The framework is illustrated in Figure~\ref{fig:design}. Specifically, 

\begin{itemize}
    \item \stage describes that visitors seek guidance at different timings of their visit, including the \textbf{beginning}, \textbf{in progress}, and the \textbf{ending}. This dimension is important because visitors' needs can shift among different stages, such as from an initial overview and familiarization in the beginning, to a more focused exploration in progress, and finally a reflection and summary at the end.
    \item \info is data or insights visitors seek, which can be  \textbf{spatial} information that helps visitors navigate and plan paths, \textbf{semantic} that deepens their understanding and appreciation of the exhibits, and \textbf{social} information that enhances the overall museum journey by fostering a sense of community. 
    \item A \task is what the guidance should facilitate visitors to achieve with the above information. The guidance is expected to facilitate \textbf{information enhancement} that provides insights to visitors, \textbf{personalized preference} that fits visitors' individual habits and choices, and \textbf{navigation} that directs visiting paths. 
    \item \response is LLM's responses to visitors' various guidance-seeking contexts.  To facilitate user tasks based on desired information in different stages, LLM can provide different forms of various feedback, including \textbf{avatar}, \textbf{voice}, \textbf{text window}, \textbf{minimap}, \textbf{signpost}, \textbf{highlight}, and \textbf{virtual screen}.
\end{itemize}

\subsection{Common Multi-modal Feedback Mechanism}
\label{ssec:context_walkthrough}

\begin{figure}[t]
    \centering
    \includegraphics[width=0.98\linewidth]{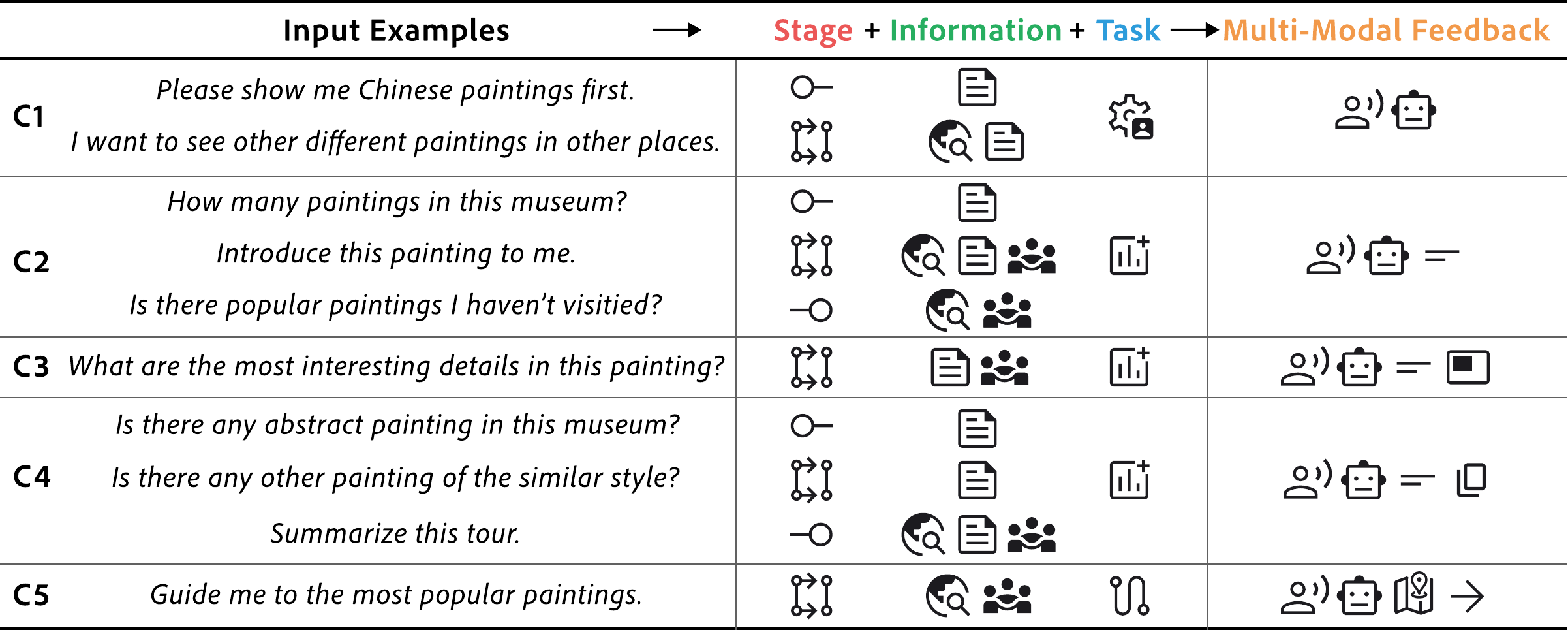}
    \caption{Five common multi-modal feedback combinations summarized from our design framework.}
    \label{fig:context}
\end{figure}

Following our design framework, we revisited the formative study results and organized mappings between guidance-seeking contexts and expected feedback.
Five common feedback combinations emerged. We list these combinations and their possible guidance-seeking contexts, together with visitors' original input examples in Figure \ref{fig:context}. 

\begin{customenum}
\item
\textbf{Voice + Avatar.}
Visitors want to specify their personalized preference (\task) with LLMs in the forms of voice and avatar (\response). They can specify their preferences regarding semantic information (\info) in the beginning (\stage) by saying \q{show me Picasso paintings first}, or regarding spatial and semantic information (\info) in progress (\stage) by requesting \q{I want to see some different paintings in other places}.

\item
\textbf{Voice + Avatar + Text Window.} 
Visitors want the combination of voice, avatar, and text window (\response) to achieve general information enhancement (\task).
For example, visitors may require semantic information (\info) at the beginning (\stage) of their visits and ask LLM \q{how many paintings are in this museum}.
They may also require spatial, semantic, and/or social information (\info) when their visit is in progress (\stage) by asking \q{introduce this painting to me}.
At the ending (\stage), they may ask \q{are there popular paintings I haven't visited} with spatial and social information (\info).

\item
\textbf{Voice + Avatar + Text Window + Highlight.}
When their visit is in progress (\stage), they may want information enhancement (\task) regarding semantic and social information (\info) for specific areas of an artwork. In this case, they want the combination of voice, avatar, text window, and highlight (\response) to provide guidance on their questions like \q{what are the most interesting details in this painting}.

\item
\textbf{Voice + Avatar + Text Window + Virtual Screen.}
When their information enhancement (\task) involves multiple exhibits in the museums, they want the combination of voice, avatar, text window, and virtual screen (\response). The virtual screen can provide semantic information (\info) to their questions like \q{is there any abstract painting in this museum} asked in the beginning (\stage) as well as questions like \q{is there any other paintings of the similar style} asked in progress (\stage). Virtual screen can also supplement voice and text window by providing spatial, semantic, and social information (\info) to requests like \q{summarize this tour} at the ending (\stage) of a visit.

\item
\textbf{Voice + Avatar + Minimap + Signpost.}
When visitors require navigation (\task) based on semantic and social information (\info) when their visits are in progress (\stage) by saying \q{Guide me to see the most popular painting}, they want the assistance of voice, avatar, minimaps, and signpost.

\end{customenum}

\section{\tool}
\label{sec:prototype}

This section introduces the design of \tool, an interactive voice-controlled system that enhances LLMs with domain-specific knowledge to improve virtual tour guidance.
The development of \tool is an iterative design process guided by design space (Sect.~\ref{sec:design-space}).
Below we provide design details on the multi-modal feedback mechanism integrated into \tool (Sect.~\ref{ssec:feedback_design}), followed by a two-stage framework that customizes LLMs to convert vague visitor requirements into multi-modal feedback for different guidance-seeking contexts (Sect.~\ref{ssec:llm}).

\subsection{Multi-modal Feedback Design}
\label{ssec:feedback_design}

\begin{figure}[t]
  \centering
  \includegraphics[width=0.99\linewidth]{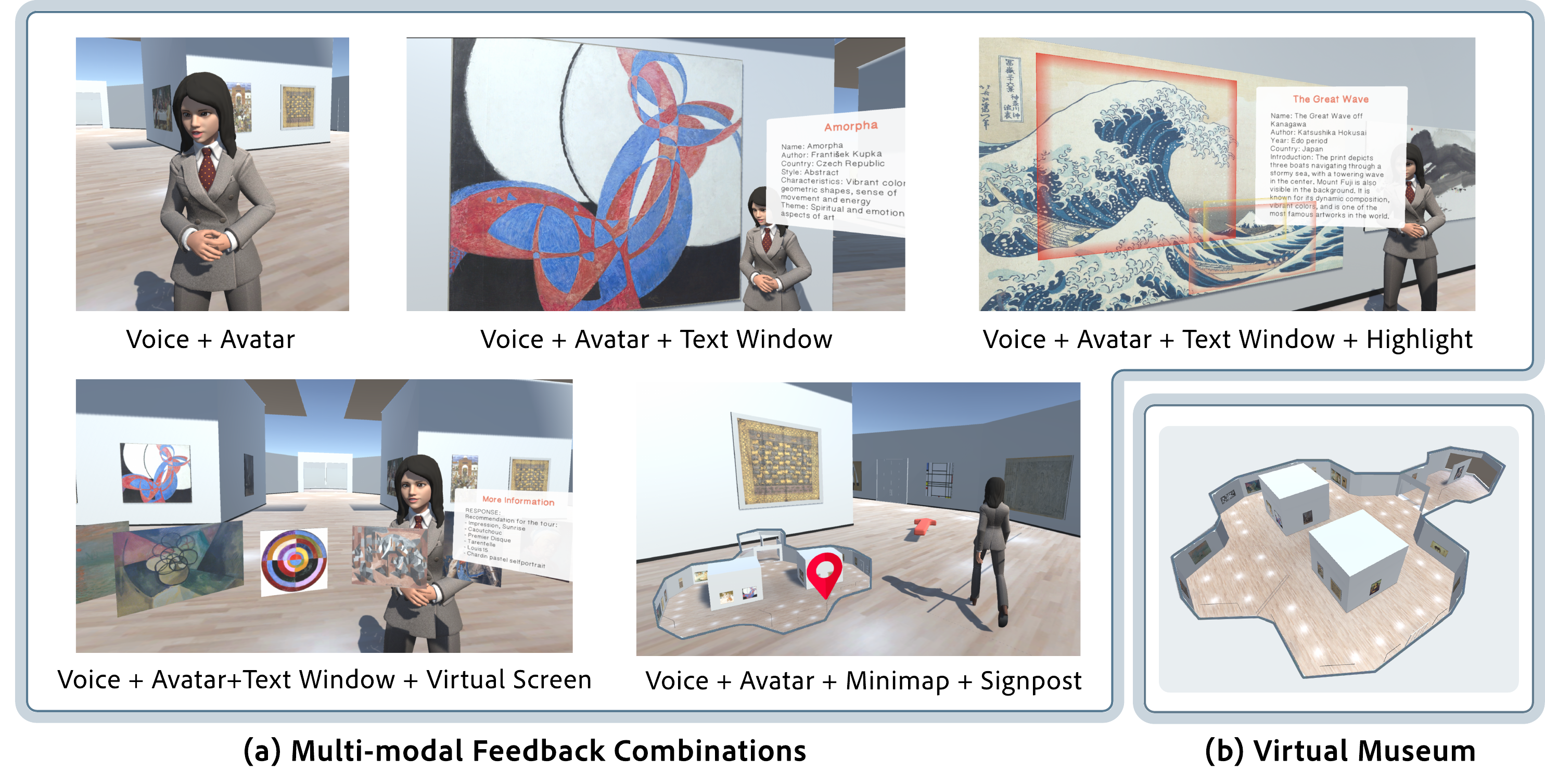}
  \vspace{-2mm}
  \caption{Implementation of \tool: (a) multi-modal feedback combinations for common guidance-seeking contexts and (b) a simulated virtual reality museum.}
  \vspace{-2mm}
  \label{fig:response}
\end{figure}

Based on the formative study and design framework, we derive a multi-modal feedback mechanism with five common feedback combinations based on different tour guidance contexts (Sect.~\ref{ssec:context_walkthrough}).
These feedback combinations serve as elementary feedback designs to support more complex touring experiences.
We collaborated closely with two museum enthusiasts to implement these feedback designs (Figure~\ref{fig:response}(a)), as follows.

\subsubsection{Voice + Avatar}
The participants indicated their preference for interacting with a tour guidance system that simulates real-world touring, including the use of voice narration for natural conversations and a virtual avatar to mimic a real guide.
As such, we have implemented \voice + \avatar (\textbf{C1}) as the basic feedback mechanism.
The \voice audibly conveys LLM responses to the visitor immediately when LLM generates answers to visitors' inquiries.
The \voice includes simple conversations like \q{what can I help you today}, as well as complex narrations such as detailed descriptions of artwork.  
This enables visitors to access unfamiliar knowledge with minimal reading effort, facilitating immersion in the virtual environment.
The \avatar includes a wide range of interactions like facial expressions and body postures aligned with the auditory content without spatial movements.
For instance, the \avatar will stand in front of the visitor when the \voice is \q{what can I help you}, or show a hand gesture pointing to the painting while standing beside it when the \voice is introduced as \q{this painting describes something}.
In many situations, \voice + \avatar feedback provides visitors with the most natural form of communication feedback without extra visual cues.

\subsubsection{Voice + Avatar + Text Window}
In the context of enhancing information access, we have determined that visitors desire a direct and efficient means to explore unfamiliar knowledge.
Such situations require the feedback design \textbf{C2} that encompasses three interaction modalities: 1) a \voice audibly responding to visitor inquiries, 2) an \avatar displaying speech animations, and 3) a \abs presenting textual responses.
Initially, we displayed the complete LLM responses and positioned them directly in front of visitors at a short distance.
However, the museum enthusiasts reported experiencing fatigue and discomfort due to the dense textual narratives.
They also found it distracting when the content was positioned just in front of the viewer, separating their line of sight from the environment. 
As a result, we revised the design by summarizing key points from the LLM responses and positioning the abstract text in the front-right of the visitor's view. 
This enables visitors to comprehend the content at a glance with minimal physical actions.
Additionally, we introduce a semi-transparent background for \abs to distinguish text from the environment while reducing visual obstruction.

\subsubsection{Voice + Avatar + Text Window + Highlight}
During a tour, visitors may encounter an artwork of interest and wish to delve deeper into its details by asking questions like \q{what are the noteworthy details in this artwork?}.
However, \textbf{C2} feedback is not sufficient for individuals lacking expertise in art.
To address this issue, we have integrated \highlight, which employs color-bordered squares on a single artwork to highlight the regions of interest determined by LLM, resulting in \textbf{C3} feedback that includes \voice, \avatar, \abs, and \highlight.
The colors of the squares are determined by the significance of the respective regions, which are more important regions highlighted in dark red colors.
The importance of each region is determined by the LLM response.
The design of \highlight also undergoes an iterative process.
Initially, we highlighted all regions of interest at all times when visitors inquired about the general information of an artwork.
However, visitors reported confusion arising from \highlight not aligning with the content conveyed by \voice and \abs.
As a result, we updated the timing of \highlight appearances by displaying it only when visitors expressed interest in detailed information about the artwork and only when \voice is discussing the corresponding regions.

\subsubsection{Voice + Avatar + Text Window + Virtual Screen}
In addition to the exploration of detailed information about individual artworks, information enhancement also encompasses another specific context related to multiple artworks.
At the beginning and the end of a tour, visitors may seek an overview or summary of multiple artworks within the virtual museum.
During the tour, visitors may also wish to compare the current artwork with others, even if those artworks are located some distance away.
\textbf{C4} feedback that consists of \voice, \avatar, \abs, and \mirror, can facilitate both the overview and comparison contexts through the incorporation of an additional \mirror.
The designs for \voice and \avatar remain consistent with the previous ones, while \abs is modified to present varying levels of information based on different stages. 
This change is in response to visitor feedback, indicating a preference for a general introduction for overviews and comprehensive details for comparisons.
Furthermore, \mirror showcases miniature representations of each artwork referred to in LLM responses, positioned slightly lower and directly in front of the visitor to prevent obstruction of their line of sight.
All artworks in \mirror are oriented toward the visitors to guarantee clear visibility, and the order of artworks displayed in \mirror corresponds to their appearance order in \abs.

\begin{figure}[t]
  \centering
  \includegraphics[width=0.99\linewidth]{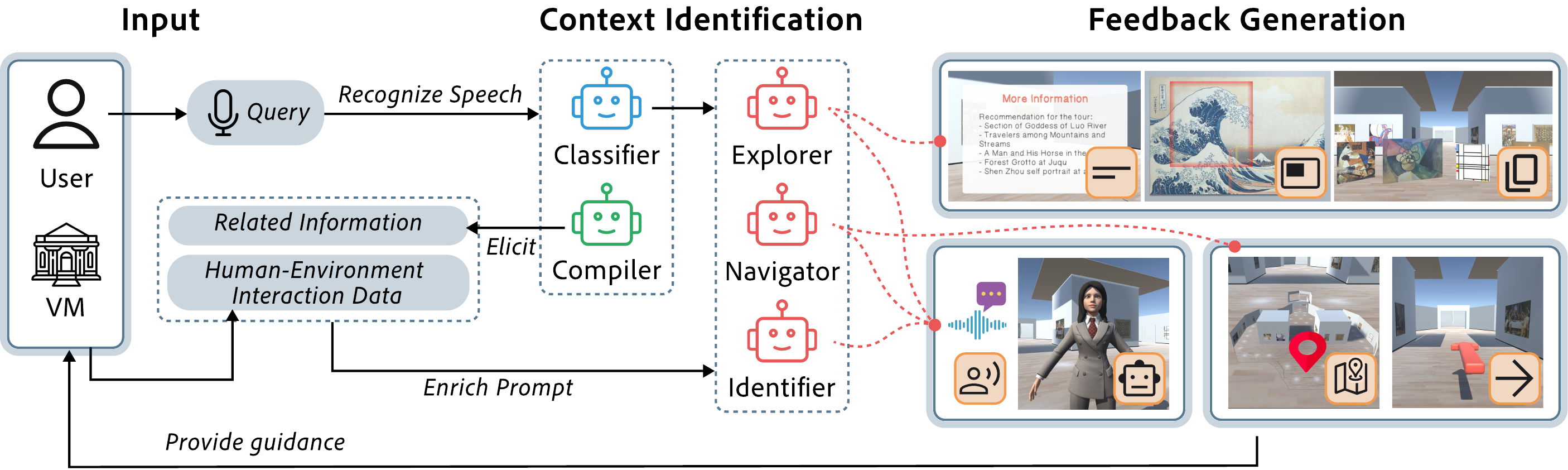}
  \caption{\tool is a voice-controlled tour guidance system with a two-stage framework: 1) \emph{context identification} stage converts visitor natural language input into various guidance-seeking contexts, and 2) \emph{feedback generation} stage generates multi-modal feedback combinations based on task-specific LLM responses.}
  \vspace{-4mm}
  \label{fig:workflow}
\end{figure}

\subsubsection{Voice + Avatar + Minimap + Signpost}
Navigating through a virtual environment differs significantly from tasks involving information enhancement and personalized preferences due to spatial movements.
The participants expressed a stronger preference for understanding their current location and destination within the virtual environment, prioritizing detailed information about the specific paintings.
Therefore, we have devised a novel multi-modal feedback combination \textbf{C5}, comprising \voice, \avatar, \map, and \sign, with the specific purpose of guiding visitors in navigating the virtual museum.
The content of \voice is tailored to provide concise navigation commands, such as \q{follow me} or \q{this way please}, rather than offering detailed introductions to specific paintings or the overall tour.
Differing from \textbf{C1-C4}, the \avatar has additional spatial movements along with facial expressions and body postures.
This movement is manifested as the \avatar walks in front of the visitor, positioned to their right at a short distance, to provide navigational guidance.
\map and \sign represent two novel feedback designs explicitly aimed at enhancing visitors' spatial awareness during their virtual tour.
The \map becomes visible only when the visitor is actively walking within the museum environment and vanishes when the visitor resumes communication with the system, identified as the completion of a navigation task.
It indicates the visitor's position on a miniature representation of the current environment using a marker.
We position the \map in the left front of the visitor to ensure uninterrupted visibility while walking, free from any obstructions.
The \sign indicates the direction from the visitor's current position to the intended destination using a directional arrow.

\subsubsection{Other Feedback.} 
\label{sssec:other-feedback}
We also add two other feedback designs to enhance the usability of our system and provide an immersive virtual museum tour experience, including:

\begin{itemize}

\item
\textbf{Visitor avatar.} To enhance embodiment and presence in the virtual reality environment, we introduce an additional visitor \textit{avatar}. Simulating face-to-face communication, the tour guide \avatar consistently maintains its orientation toward the visitor \textit{avatar}. 
Besides, the visitor \textit{avatar}'s spatial movement is automated, following the tour guide \avatar without the need for visitors to manually control movement using VR controllers. 
The speeds of both avatars are preset as a constant to mimic a comfortable movement.
Although manual navigation offers flexibility, our primary objective of such interaction design is to replicate real-world scenarios by establishing a hands-free experimental environment.

\item
\textbf{Conversational interface.} To improve visitor comprehension of the voice input modality, we have designed a conversational interface for visitors to observe their input inquiries. 
This interface will automatically disappear once our system provides multi-modal feedback.

\end{itemize}

\begin{figure}[t]
  \centering
  \includegraphics[width=0.98\linewidth]{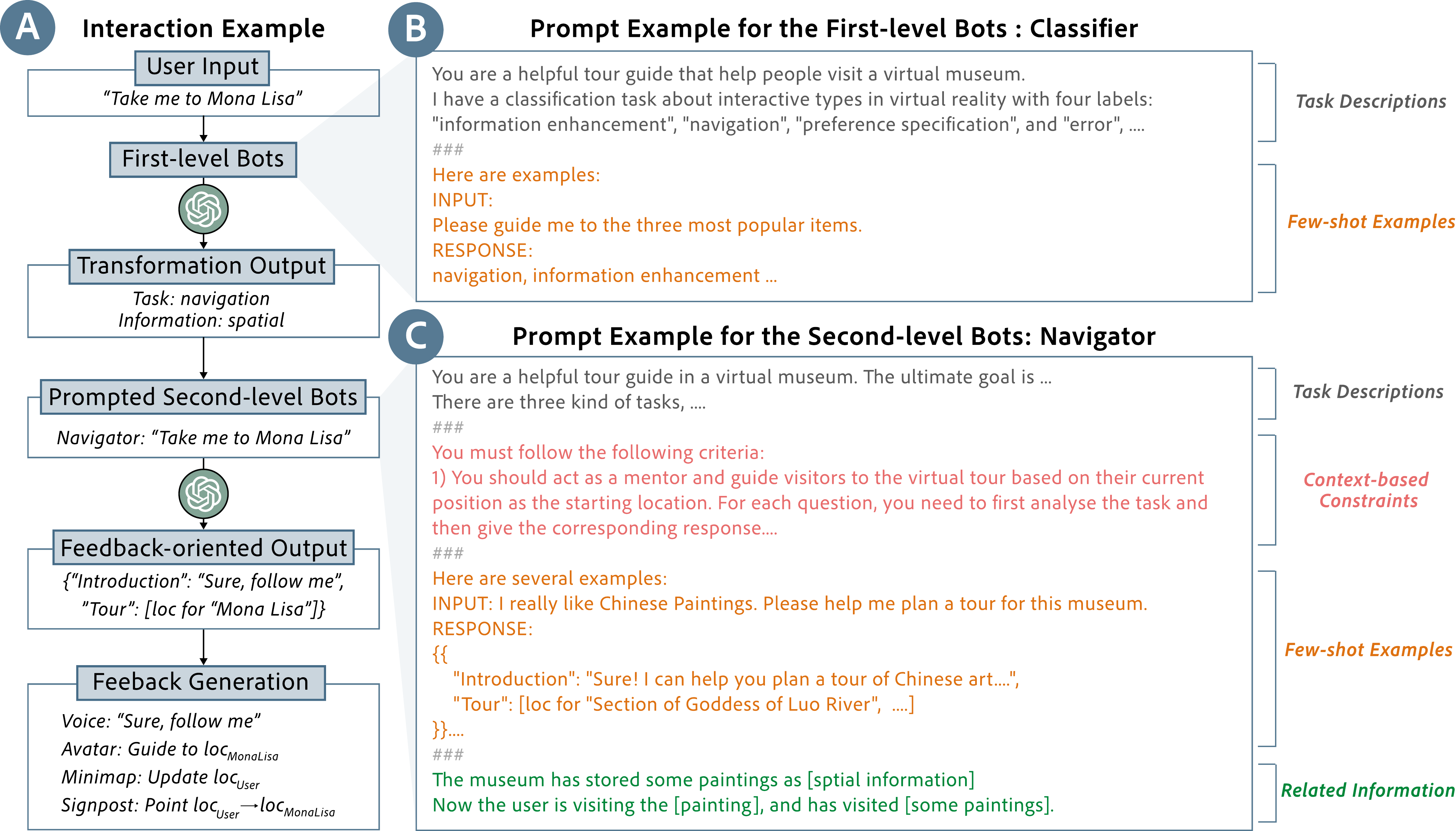}
  \caption{Examples for (A) inputs and outputs of our two-stage pack-of-bots strategy and (B) prompt techniques of the first-level bots and (C) the second-level bots.}
  \vspace{-2mm}
  \label{fig:LLM}
\end{figure}

\subsection{LLM-based Feedback Generation}
\label{ssec:llm}
We develop a two-stage framework (Figure~\ref{fig:workflow}) that first converts visitors' natural language inputs into diverse guidance-seeking contexts and then generates multi-modal feedback.
The framework harnesses a pack-of-bots strategy with multiple LLM-based bots tailored to specific roles and tasks, to meet the requirement for interactive feedback at a responsive rate.
Each LLM-based bot necessitates well-designed prompts to elicit reasonable and accurate responses.
In the following, we introduce how we craft prompts carefully in order to align with our design framework and generate multi-modal feedback, \ie, prompt engineering.
Prompt engineering structures the input provided to the model to enable more tailored responses for specific tasks or objectives instead of fine-tuning the entire model.
All of our prompted chatbots follow a few-shot learning paradigm.
In practice, we utilize GPT-3.5 as the model to be prompted.
We provide some examples of the two-level pack-of-bots strategy and prompt techniques of each level of bots in Figure~\ref{fig:LLM}. 
The designed prompts for each bot are presented in supplementary materials.

\subsubsection{Stage 1: Context Identification.} 

The first stage includes \textit{Classifier} and \textit{Compiler} bots.
When visitors input their inquiries via voice, \tool first converts the spoken words into textual natural language.
\textit{Classifier} translates the visitor's intent into one or more of the interaction \tasks, while \textit{Compiler} identifies the types of \info required to support the second-stage bots in completing their tasks.
Based on the task classification results, \textit{Classifier} directs visitor inquiries to the respective task-specific bot, namely, \textit{Explorer} for information enhancement, \textit{Identifier} for preference specification, and \textit{Navigator} for navigation.
It is important to note that the task-specific prompts used by the second-stage bots are dynamically enriched with contextual \info obtained through the \textit{Compiler} and human-environment interaction data.

Both \textit{Classifier} and \textit{Compiler} effectively perform a multi-classification task. 
Therefore, we craft a straightforward prompt template for task description and provided a few-shot examples, as illustrated in Figure~\ref{fig:LLM}(B).
The task description part encompasses perspective-tasking prompts (\eg, \q{You are a museum tour guide}), concrete definitions of different categories (\eg, \q{information enhancement refers to...}), and task specification prompts (\eg, \q{Please classify visitor intent into these categories}).
The few-shot examples include various visitor inputs paired with corresponding results. 
For instance, an input such as \q{guide me to the most popular paintings} falls under both information enhancement (\ie, identifying the most popular painting) and navigation (\ie, guiding the visitor to this painting) tasks.

\subsubsection{Stage 2: Feedback Generation.} 
The second stage consists of three bots, each prompted with completing distinct \tasks and delivering task-oriented responses to facilitate the generation of multi-modal \response combinations, as outlined in the aforementioned design framework across various \stages.
In addition to providing the standard guidance for \voice and \avatar, \textit{Explorer} can extract key points from full introductions for \abs, detect indications of interest in specific artworks for \highlight, and identify multiple mentioned artworks within the introduction for \mirror.
\textit{Navigator} provides brief navigational commands for \voice and presents a list of ordered artworks to establish a touring path for \avatar, \map, and \sign.
\textit{Identifier} exclusively focuses on natural communication for \voice and \avatar interactions.
While \tool may classify a single visitor input into two or three distinct tasks, we assume that visitors expect to get more information for a single \task but introducing more feedback combinations together for different \tasks could potentially confuse our audience.
Therefore, we prioritize guidance feedback combinations based on a hierarchy that navigation is given precedence over information enhancement, which takes precedence over personalized preference.

The three bots are specifically designed to facilitate the creation of multi-modal feedback combinations. 
Consequently, their prompt templates go beyond task descriptions and few-shot examples.
The templates also incorporate additional context-based constraints and related information components that are dynamically updated based on \textit{Compiler}'s responses (Figure~\ref{fig:LLM}(C)).
Context-based constraints encompass task-specific criteria (\eg, \q{The tour should be novel and interesting}), stage differentiation (\eg, \q{Please provide answers in less than 4 sentences if the visitor is in the middle of the tour}), and response format specifications (\eg, \q{Please format the tour response in JSON format, including a brief introduction and a list of paintings}).
These additions enhance the responses' robustness to support feedback generation throughout all stages of the virtual tour.
Related information includes contextual details recommended by \textit{Compiler} (\eg, spatial positions or semantic introductions for each painting) and human-environment data (\eg, the currently visiting painting or tour history). 
This supplementary information enriches the responses generated by the bots, making them more contextually relevant and informative.

We provide a comprehensive interaction procedure example in Figure~\ref{fig:LLM}(A) that illustrates the entire process from visitor input to multi-modal feedback, to enhance understanding of the workflow.
When a visitor inputs \q{take me to Mona Lisa}, \textit{Classifier} identifies this as a \q{navigation} task, while \textit{Compiler} recognizes that the second-stage bots require additional \q{spatial} information, specifically the location of the Mona Lisa. 
Consequently, this location information is incorporated into the prompt of \textit{Navigator}, along with the visitor's current position.
\textit{Navigator} responds to navigational commands as well as a tour list specifying the location of the Mona Lisa in a JSON format following the predefined constraints.
Given the specified location, the \textit{avatar}'s navigation path from its current position to the destination is determined as the shortest path within its visual range, integrating collision avoidance mechanisms for efficient and safe movement.
As a result, the visitor ultimately receives multi-modal feedback, comprising an \avatar saying \q{Sure! Follow me.} and guiding the visitor to the Mona Lisa, a \map displaying the visitor's changing location within the museum, and a \sign indicating the direction from the visitor's location to the Mona Lisa's location.

\section{Example Cases}
\label{sec:case}

\begin{figure}[t]
    \centering
    \includegraphics[width=0.99\linewidth]{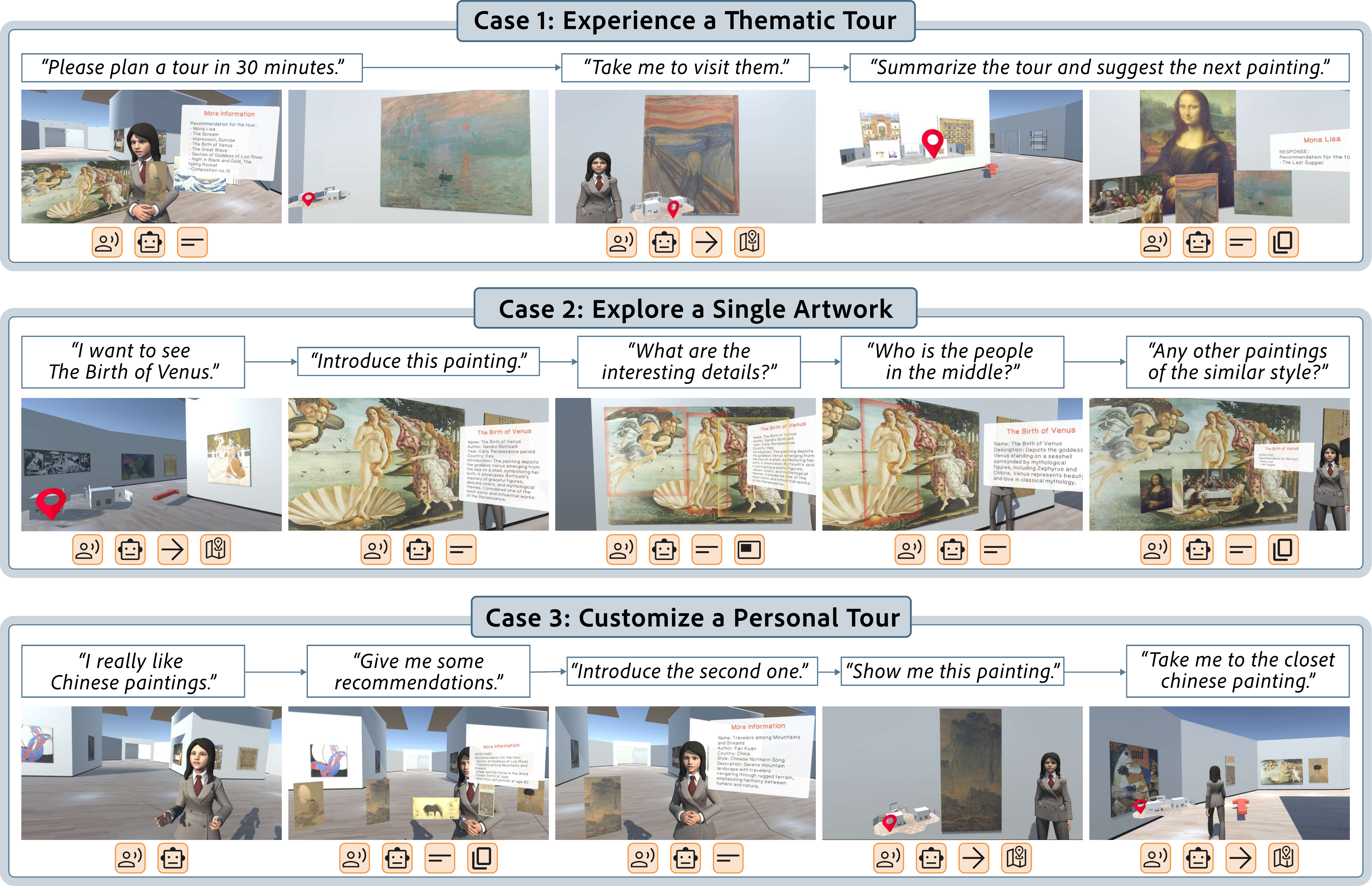}
    \caption{Three example cases for virtual museum tour guidance experience with our system \tool: experience a thematic tour, explore a single artwork, and customize a personal tour.}
    \vspace{-2mm}
    \label{fig:cases}
\end{figure}

\textbf{Virtual Museum Simulation.}
Our study focuses on art museums for designing LLM-based interactions, reflecting the predominantly solitary nature of art museum experiences where visitors engage in personal reflection and appreciation of artworks~\cite{hornecker2019human}.
To simulate this environment, we created a virtual art museum, modeled after a real-world museum's single-floor layout with two interconnected sub-spaces. 
Due to limited access to original artworks, our exhibit featured 35 renowned paintings spanning various styles and periods, from French Renaissance to Chinese ink paintings, to ensure a broad and inclusive representation of art collections for evaluating our system's effectiveness in an immersive virtual setting.
The entire system is implemented using Unity3D and Oculus Quest 2. 
The engineering of large language models is conducted using Python in conjunction with Flask for data transportation between Python and Unity3D.
We preload contextual information into our system, such as the names and positions of the paintings.
We explore three common cases identified from the user study (shown in Figure~\ref{fig:cases}): (1) experience a thematic tour, (2) explore a single art work, and (3) customize a personal tour.
Appendix~\ref{sec:appendix} presents illustrative examples of questions and answers generated by \tool across these cases.
These examples serve to demonstrate the system's capabilities in addressing diverse user inquiries within the context of virtual museum tours.

\textbf{Case 1: Experience a Thematic Tour.} 
This case illustrates a typical scenario where \tool facilitates a thematic tour experience. 
Initially, a user has no idea about what to visit in this museum and asks \tool to suggest a tour achievable within 30 minutes (\eg, \textit{"Help me plan a tour in 30 minutes"}).
\tool presents eight of the most popular paintings within the museum, utilizing both \abs and \. 
The \avatar accompanies this display with \voice assistance.
Then the user expresses a desire to visit the first three recommended paintings, namely, \textit{"Mona Lisa"}, \textit{"Last Supper"}, and \textit{"The Scream"}, in sequence (\eg, \textit{"I want to see the first three paintings one by one"}).
Our tool guides the user through a thematic tour, efficiently navigating the shortest path with the assistance of \sign and \map to indicate spatial movements.
Upon concluding this initial tour, the user finds available time and expresses an interest in further exploration. 
They request \tool to summarize their tour and provide additional recommendations (\eg, \textit{"Summarize this tour and give me some suggestions"}). 
Considering the user preference for famous artworks, \tool proceeds to suggest a previously unvisited popular painting, \textit{"Impression, Sunrise"}.

\textbf{Case 2: Explore a Single Artwork.}
This case demonstrates a scenario in which the user explores their preferred artworks. 
When a user has decided to visit the painting \textit{"The Birth of Venus"}, s/he initially requests \tool guidance to this artwork. 
Upon arriving in front of the painting, the user seeks a basic understanding of it.
Given \tool's knowledge from the previous navigation task for this painting, it provides a brief introduction, including the painting's name, author, year, and a one-sentence description of its content.
With a desire to delve deeper, the user asks \tool for more meaningful details about \textit{"The Birth of Venus"} (\eg, \textit{"What are interesting details in this painting"}), \tool responds by highlighting three key figures in the painting, offering both auditory and textual information.
The user's curiosity leads to inquiry about the central figure (\eg, \textit{"Who is the person in the middle"}) revealed by \tool to be the Goddess Venus.
After completing their visit to this single artwork, the user expresses an interest in discovering more paintings of a similar style (\eg, \textit{"Is there any other similar paintings"}). 
\tool responds by presenting several paintings of the oil painting style in \mirror.

\textbf{Case 3: Customize a Personal Tour.}
This case demonstrates how a user customizes a tour based on their personal interests using our tool. 
The user first expresses a specific preference for Chinese paintings, stating, \textit{"I really like Chinese paintings"}. 
Throughout the subsequent conversation, \tool remembers this personal preference for the user and customizes guidance based on various personalized preferences.
When the user continues to request \textit{"some recommendations"} without specifying a painting style, \tool consistently suggests five Chinese paintings. 
Then the user selects their most preferred painting from \mirror and then asks \tool to provide more information (\eg, \textit{"Introduce the second one"}) and guide them to visit the chosen painting (\eg,  \textit{"Show me this painting"}).
After exploring this individual Chinese painting, the user still wishes to visit more Chinese paintings.
They request to continue the tour (\eg, \textit{"Take me to the next painting"})  and \tool navigates them to the nearest Chinese painting.

\section{User Study}
\label{sec:userstudy}
We conducted a user study to evaluate the usefulness and engagement of LLM-enhanced multi-modal feedback designs and the overall virtual tour experience of the whole \tool.

\subsection{Participants}
We recruited 12 participants (P1-12; M = 6, F = 6; Age: 22 - 39) via the university mailing list and gathered information about their prior VR experience and museum-visiting preferences. 
All participants had normal or corrected-to-normal vision.
Their VR experience varied: 2 participants had more than 20 hours of experience, 3 had 10-20 hours, 5 had less than 10 hours, and 2 had never used VR.
Most participants expressed a strong likelihood or likelihood of visiting museums, with only 2 participants indicating a neutral attitude.
We also gathered detailed demographic data to encompass participants' non-exclusive interests in different types of museums and art, along with their motivations for engaging in a virtual museum environment.
Participants indicated a diverse range of interests: 75\% showed a preference for history museums, 58.3\% for art museums, and 41.7\% participants expressed interest in cultural and science museums, 
Regarding art styles, a majority (58.3\%) were interested in classical art, while both contemporary and modern art engaged the attention of 41.7\% of participants.
Motivations for visiting virtual museums were categorized in accordance with Falk and Dierking's framework~\cite{falk2013museum}. 
The predominant motivation was exploration, with 75\% of participants identifying as explorers, motivated by curiosity. 
Other significant motivational categories included professionals and hobbyists deepening their knowledge (41.7\%), experience-seekers looking for a special experience (33.3\%), facilitators who assist the experience of others (16.7\%), and those seeking a restorative or contemplative experience (8.3\%).
Each participant received a \$10 gift card as compensation.

\subsection{Design and Procedures}
This study contains four sessions as follows: (1) a tutorial session for familiar participants with VR and the tool, (2) a task completion session to provide experiences of different feedback designs, (3) a free exploration session to simulate a complete real-world guided tour experience, and (4) a post-study interview to collect user qualitative feedback.
Throughout the experiment, participants had the liberty to pause the study and take breaks whenever desired. 
The entire experiment had an average duration of 90 minutes.

\textit{Tutorial.} 
The first session serves as a tutorial to acquaint participants with VR and the operational procedures of the tool.
We started the experiment with a 10-minute introduction about the experimental procedure and design framework.
VR headset was adjusted to ensure the sample text was clearly visible in front of them.
We then provided a set-up training on how to use VR and utilize our tool through voice commands.

\begin{figure}[t]
    \centering
    \includegraphics[width=0.99\linewidth]{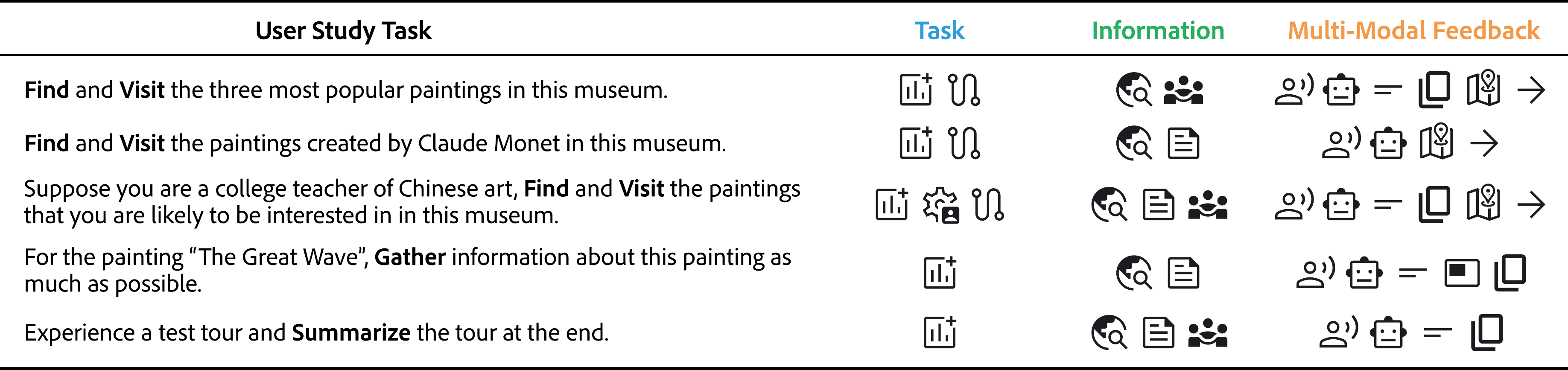}
    \caption{Tasks designed for the user study to make participants experience various contexts and multi-modal feedback combinations.}
    \label{fig:study-task}
\end{figure}

\textit{Task Completion.} 
The second session is a task completion session specifically designed to offer an experience of various feedback designs.
It focused on the effectiveness and usefulness of our design framework and explored user behavior for how to use each feedback design or their combination.
Participants went through several tasks illustrated in Figure ~\ref{fig:study-task}, aiming to simulate common guidance-seeking contexts and experience all primary multi-modal feedback.
We encouraged participants to interact with the tool naturally instead of completing the task as quickly as possible.
The participants were then asked to rate and provide feedback on the functionality of each feedback design.

\textit{Free Exploration.} 
The third session is a free exploration and follows participants' familiarization with the LLM-enhanced multi-modal feedback combinations.
This session is focused on the engagement and usability of our entire prototype.
Participants were asked to complete a museum tour with our tool, simulating a real-world scenario.
The tour encompassed designing a personalized interest-based tour beforehand, navigating the museum, visiting points of interest, and summarizing the overall tour.
We asked them to rate the overall user experience of the system and to offer insights into their interactions and conduct during the guided tour experiences.

\textit{Post-study Interview.} 
The last session is a post-study evaluation. 
We employed a semi-structured interview format to gather qualitative feedback from participants regarding their utilization of the LLM in tour guidance.

\subsection{Measures}
\textbf{Part 1.} 
For each feedback design, participants were asked to fill in a 7-point Likert scale questionnaire (ranging from low to high), aimed at gathering subjective ratings in terms of such as understandability (\textit{"it was easy to learn"}), focused attention (\textit{"it was focused"}), usefulness (\textit{"it was helpful"}), interest (\textit{"it was fun to use"}), and novelty (\textit{"it was novel"}). 
Additionally, we collected oral feedback from participants, exploring the reasons behind their preference for different feedback designs and offering suggestions for potential design enhancements.

\noindent
\textbf{Part 2.} 
We also collected their 7-point Likert scale ratings for the engagement and usability of the \tool system as a whole.
We formulated questions to evaluate four aspects of engagement from prior work~\cite{o2010development}, including feedback (\textit{"it was helpful"}), interest (\textit{"it was fun to use"}), in control (\textit{"it was in control"}), and motivation (\textit{I was likely to use"}).
We also measured the usability with four questions tailored from our main target, like \textit{"it was easy to use"}, \textit{"it was easy to learn"}, \textit{"it was natural to use"}, and \textit{"it was tailored to my interest"}.
In order to compare simulation experience in virtual reality to the real-world scenarios, we also collected user experience regarding virtual tours~\cite{bowman1998methodology} through a 7-point Likert scale, encompassing user comfort (\textit{"it was comfortable"}), presence (\textit{"I felt immersed or 'being within' the environment}), and spatial awareness (\textit{"it was easy to know where I was"}).

\noindent
\textbf{Part 3.}
During the post-study interview, we gathered qualitative feedback from participants regarding their interactions with various multi-modal feedback designs. 
They also provided insights into the strengths and weaknesses of using the tool in real-world scenarios, as well as their expectations for tour guidance enhanced by LLM capabilities.

\begin{figure}[t]
    \centering
    \includegraphics[width=0.99\linewidth]{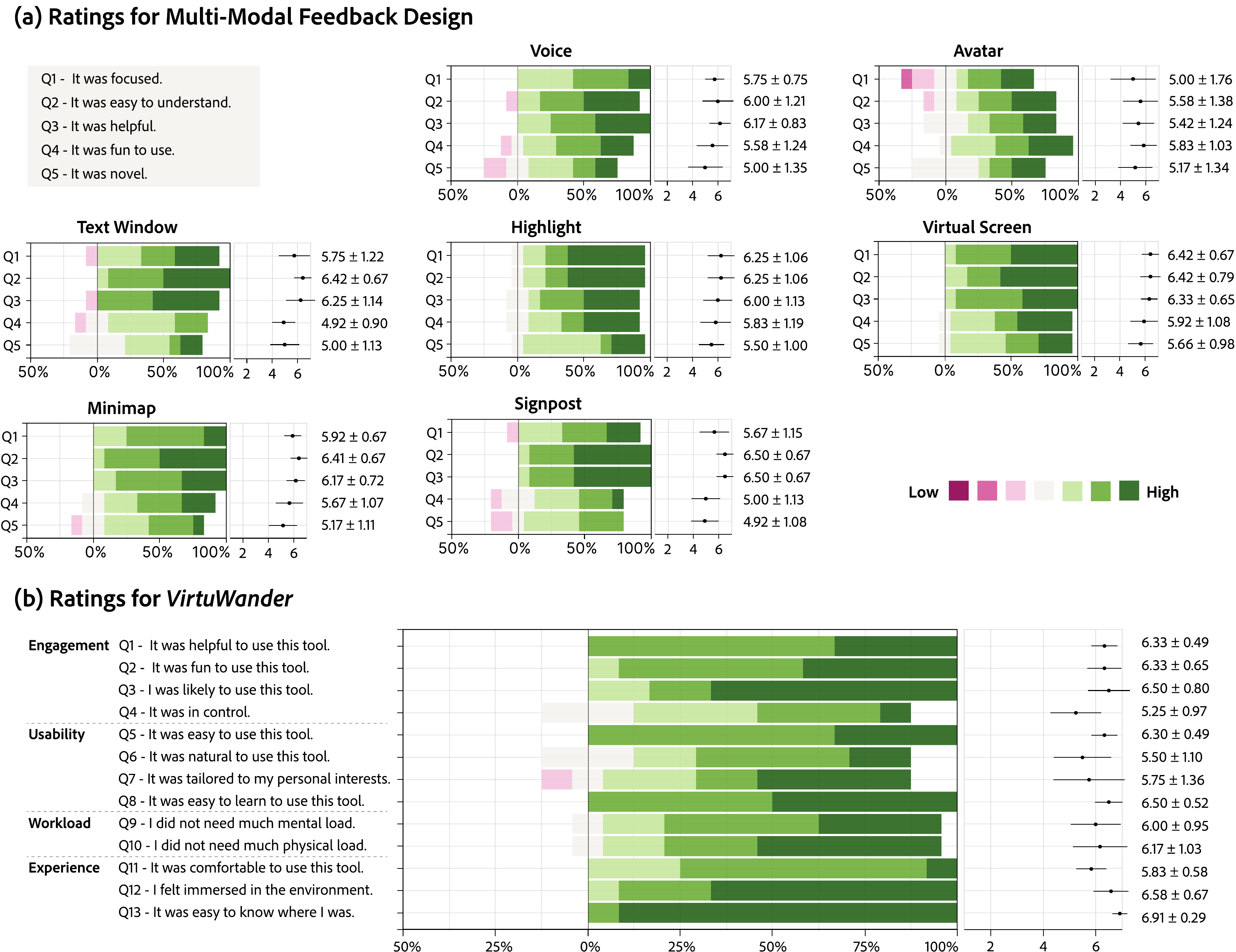}
    \caption{Quantitative user study results. Participants (a) rated the engagement for all feedback designs and (b) confirmed that \tool enhanced the overall virtual tour guidance experience. Each subplot is composed of a horizontal bar chart to show the percentages of each rating score (left) and a box plot showing mean and standard deviation values (right).}
    \label{fig:rating}
\end{figure}

\subsection{Results \& Findings}

\subsubsection{How do visitors interact with multi-modal feedback in virtual museum tours?}

Figure~\ref{fig:rating} shows the ratings for our multi-modal feedback design. Overall, the ratings were positive. In the following, we will report how our participants interacted with each type of feedback.

\textbf{Voice }
Voice received four points or above from all participants for Q1 and Q3, indicating it was useful and helpful. 
It is not surprising that some participants felt voice was not fun (Q4, N=2) and novel (Q5, N=4), because it was the most common interaction modality in daily life. 
When voice failed to provide key information, one participant felt not easy to understand voice (Q2) and suggested, \q{I always lose important points in long sentences} (P8).

\textbf{Avatar }
Avatar has rated four points or above from all participants for Q1, Q4, and Q3, indicating that it is helpful, fun to use, and novel. 
Participants particularly thought the avatar was fun to use. 
For example, P8 said, \q{Avatar provides a sense of companion, though, it does not provide meaningful information in the tour}. 
Participants suggested the avatar appearance to further improve its fun. 
Specifically, a majority of participants (N=5) suggested a cute and miniature cartoon avatar to improve a sense of affinity. 
However, 4 participants gave low ratings to Q1 because avatars distracted them. 
They wished that \q{the avatar only appears when needed} (P10) for free-roaming or preventing view blocking.

\textbf{Text Window }
All participants rated four points or above for Q2, indicating they felt easy to understand text window. Surprisingly, all participants also rated four points or above for Q5, feeling the text window novel. 
This was because the text window was displayed near eyes when requiring information, which was currently \q{not feasible in real worlds} (P10).
One participant felt burdensome and distractive and rated low for Q1 because sometimes the text was \q{too long and too detailed} (P3).
Similar reasons caused one low rating regarding helpfulness for Q3.
Some participants suggested improvement by matching the context and vocabulary used in text window with their personal knowledge level, such as professional or simple. 
For example, P11 commented, \q{I want to understand the text within my knowledge system}.

\textbf{Highlight }
The design of the highlight received four points or above from all participants across all five questions, indicating that participants liked to interact with it. 
Some participants thought the highlight was particularly helpful for \q{people unfamiliar with art} (P2) because it could \q{guide visitors where to look at} (P5).
We also heard different opinions from participants with art backgrounds.
They preferred highlight to \q{appear only when necessary} (P10) and to \q{customize the highlighted areas based on own interests} (P11).

\textbf{Virtual Screen }
Virtual screen received five points or above from all participants across all five questions, becoming the most popular design.
Participants liked its feature of extensive ability for searching tasks as P1 noted, \q{[Screen] build an overview of the tour, leading the process from vague searching to precise searching}.
Meanwhile, participants expected the virtual screen to offer enhanced interaction through gestures such as \q{zooming, selecting, and dragging} (P2) while presenting information in a user-friendly and non-obtrusive manner.

\textbf{Minimap }
All participants rated four points or above for Q1 to Q4 while three participants felt it was not novel. 
Some participants (N=3) suggested that minimap gave an overview of the entire space, as P5 mentioned, \q{It is clear at one glance}.
Moreover, a 3D minimap \q{improves spatial awareness and gives a sense of distance} (P6) in virtual reality particularly when the environment is complex.
To make it more helpful, participants expected the minimap to display additional spatial information, such as \q{the location of interested paintings} (P9) or \q{tour history records} (P1).

\textbf{Signpost }
Signpost received the highest average scores in terms of both ease of use and helpfulness because all participants rated five points or above. 
Signpost is very useful to \q{give a hint of want to do} (P9) and \q{provide detailed information} (P1).
However, some participants rated low for Q1 (N=1) and Q4 (N=4), because the presence of the signpost made them \q{feel less free and under pressure} (P11) when they wanted to experience free exploration.
Signpost was rated as the least novel design (Q5) as \q{it is too common} (P2).

\textbf{Combination \& Comparison }
We now reported participants' comparisons among different types of feedback.
First, when multiple feedbacks were presented simultaneously, some participants instinctively tended to prioritize voice information. For example, P1 and P7 showed a preference for listening to voice than reading text. 
However, when the voice provided excessive or irrelevant information during the tour, participants felt that \q{the process of appreciating artwork and contemplation was interrupted} (P8). 
Second, participants mentioned that highlights should appear with the text description simultaneously.
Otherwise, they might \q{fail to comprehend the purposes} (P7) of highlights. 
Furthermore, P9 suggested incorporating storytelling techniques in the text to present highlights, enhancing readability and engagement.
Third, the virtual screen received the highest average scores among all designs regarding focus (Q1), fun (Q4), and novelty (Q5).
It received high praise from participants, such as \q{more intuitive than text} (P3), \q{convenient in comparison} (P4), \q{[I can] get a lot of information in a moment} (P8), and \q{[screen] saves time} (P6).

\subsubsection{How does VirtuWander enhance virtual tour experiences?}

Overall, \tool received high ratings (Figure~\ref{fig:rating}(b)) for enhancing virtual tour experience. 
Below, we first report common user strategies during exploration. Next, we summarize feedback on how \tool improved \emph{communication, usefulness}, and \emph{immersion} in tour experience.

\textbf{Participants used various exploratory strategies with \tool.}
We have identified three primary ways in which participants initiate a tour with \tool. 
The most used strategy is to ask the system to directly guide them to individual paintings within a collection of interest, where they proceed to appreciate each artwork individually before delving into in-depth exploration.
Another frequently utilized approach begins with participants obtaining an overview of the museum's artworks. 
Then they select the interested ones and request our system's assistance in planning a thematic tour centered around their chosen pieces.
A noteworthy strategy adopted by some participants involves beginning their exploration by acknowledging statistical data results.
For instance, P3 asked about specific counts of certain painting categories, \q{How many paintings by Da Vinci are in this museum?}
Additionally, a few participants oriented their virtual museum exploration around spatial considerations, such as \q{What is the painting on the left} (P4).

\textbf{\tool supports natural and tailored communications.}
Most participants agreed that our system provides \q{a feeling of natural communication through real-time feedback} (P11) and is tailored to individual interests during guided tours.
Some participants (N=3) perceived the system's capabilities as comparable to or even superior to those of a real human guide.
It serves as \q{a personalized tour guide} (P9) for each individual and is even \q{more interactive than pre-designed voice-overs} (P5).
Many participants (N=5) acknowledged the social advantages of our system. 
Notably, it \q{reduces the pressure to communicate with a virtual avatar} (P10) and enables them to \q{interrupt the tour guide impolitely without the burden} (P5) when needed.
P11 also underscores an enhanced sense of security as \q{I will not feel overly anxious or nervous in unfamiliar environments because I can easily access unknown information}.
Nevertheless, two participants observe that engaging with our system sometimes leads to \q{easily neglecting interactions with peers} (P6), particularly in scenarios where multiple individuals are visiting together.
One participant expressed an expectation for communication to be \q{more emotionally engaging} (P4) in terms of both response content and voice.

\textbf{\tool is useful for knowledge enrichment and decision-making.}
All participants confirmed the usefulness and effectiveness of our system.
They noted that our system supports \q{a much deeper exploration for artworks} (P3) compared to traditional tour guides.
Participants also indicated that the multi-modal feedback design in our tour guidance system is \q{more intuitive for comprehending information conveyed within this environment} (P10).
Most participants (N=8) mentioned that our system can provide unknown knowledge tailored to specific needs.
P1 praised the system for its wealth of professional knowledge as \q{even as an art teacher, [the tool] could provide information beyond my expertise}.
One participant (P4) expressed expectations for enhancing our system through network searching to deliver \q{the latest information}, such as updates on recent museum exhibitions. 
Furthermore, the majority of participants (N=7) also suggested that our system helps them \q{make decisions with reduced time and human effort} (P15).
They found our system highly advantageous for \q{planning tours with time constraints in unfamiliar environments} (P12).
P4 stated the preference for allowing the tool to \q{plan the tour} instead of doing so themselves.
Additionally, participant P7 specifically favored the system's summarization feature, noting that \q{this summary report can guide me when visiting the same location next time}.

\textbf{\tool enhances engaging and immersive virtual tour experiences.}
Overall, most participants agreed that the tool enhances engaging and immersive virtual tour experiences, and they also expressed a likelihood of using our system in the future.
Many participants (N=5) indicated that \tool liberates them from electronic devices, thereby enhancing their engagement during virtual tours.
They appreciate the convenience of requesting guidance directly through voice commands, remarking that it allows them to \q{search for information without needing to pick up phones} (P6).
By disconnecting from their digital devices, they reported feeling more deeply engaged while exploring and perceiving a new environment.
P12 described it as \q{checking my phone used to disconnect me from the touring process, but now it feels like I am walking with someone who knows the place and enjoys it, which is reassuring}.
From a distinct perspective, P11 commented that our tool becomes \q{an interesting component of virtual tour experience} rather than just a tour guide assistant.
Our tool enriches the exploratory experience by \q{breaking inertia and enabling to experiment with new modes of communication} (P11).
However, some participants raised concerns about the timing of feedback appearances, noting that untimely interruptions can impact immersion. 
They expressed a desire for guidance to appear \q{at the appropriate moment} (P9), avoiding disruptions to the ongoing tour experience \q{due to feedback that appears out of nowhere} (P6).

\section{Discussion \& Future Work}
\label{sec:discussion}

Based on our observations throughout the study, we summarize some design implications for future tour guidance interactions and discuss the possibilities and challenges of extending our system to real-world scenarios.

\subsection{Implications for Future Tour Guidance}

\textbf{Enrich expressive input interaction modalities.}
\tool utilizes voice as the natural language input modality and incorporates multiple output modalities to enrich the tour guidance experience. 
However, our participants have expressed their desire for more robust input modalities to enhance their connection with both the environment and the feedback they receive.
First, it is difficult for users to access information if they can not describe artworks solely through verbal descriptions.
Additionally, as the richness of multi-modal feedback increases, users' expectations for continuous interaction with these feedback mechanisms also grow.
For instance, some participants expressed a desire to \q{physically interact with the artwork by touching it and adjusting its angle and size through gestures} (P4), as exemplified by the presented \mirror. 
Therefore, augmenting voice inputs with mid-air gestures enhances the flexibility and controllability of the tour guidance experience.
We suggest exploring additional input interaction modalities to cater to users‘ continuous interactive needs in virtual tour experiences.

\noindent
\textbf{Combine active and passive feedback modalities.}
Our system follows a passive guidance experience design, where guidance feedback is activated when users explicitly vocalize their requirements.
We believe voice commands offer the most direct and effective means to convey the user's intention with overwhelming choices.
Participants expressed concerns regarding their uncertainty about which questions to ask when they lacked familiarity with our system. 
Additionally, they occasionally found the sudden appearance of multi-modal feedback to be \q{distractive} (P12). 
As a result, we propose that an LLM-infused system should offer an alternative user experience that actively seeks user preferences and objectives while also providing guidance on how to input requirements and when to expect feedback.
To this end, we recommend a more flexible interaction approach that allows users to configure their preferred guidance styles as either passive or active. 
Future work can explore interaction design space to explore more attractive combinations of passive and active guidance feedback, as well as their timing of appearance. 

\noindent
\textbf{Support both natural and directive communication styles.}
While some participants appreciated natural communication resembling interactions with a live human tour guide, others preferred issuing direct commands to avoid superfluous conversation without concerns of impoliteness toward a virtual avatar.
In a specific environment, context-based tasks are always consistently well-defined such as visiting artworks in museums.
Therefore, it is preferable to \q{convey more with fewer words} (P3). 
Participants also expressed expectations of LLM-enhanced guidance being more intelligent in encouraging \q{more in-depth inspirations} (P7).
One idea to enhance the system's ability to discern user intent from short context-related commands is to leverage reinforcement learning with human feedback to optimize system performance.
By learning from user interaction behaviors, this technique can tailor the system responses to harmonize with both natural and directive communication styles, accommodating a broader range of user preferences.

\noindent
\textbf{Customize information magnitude and granularity.}
In human-LLM communication, the quantity and depth of information delivered by the agent should be tailored to both the user and the contextual factors. 
First, excessive knowledge can place a substantial cognitive load on users, as P10 mentioned, \q{I can only accept a limited amount of information}.
Users exhibit varying expectations, seeking detailed information in specific instances while desiring broader insights in others, which is a challenge for our system to distinguish between these nuanced preferences.
Second, the focus of information required varies across different visitor groups. 
Our system is mainly designed for general users lacking professional knowledge in the field of art.
However, some participants expressed confusion with the content and style of information furnished by our system, supposing it inconsistent with their existing knowledge framework, \q{I expect explaining professional terminology in a more joker way} (P2).
To customize the magnitude and granularity of provided information, one viable approach is to categorize visitors into distinct groups and specify more intricate roles and tasks for LLMs.

\noindent
\textbf{Ensure information accuracy and standardization.}
Our system archives the essential information about displayed artifacts and leverages the LLM's expansive knowledge base to tailor responses according to user intent. 
Despite this, some participants expressed their \q{low trust in unverifiable AI-generated knowledge} (P1) and apprehensions about \q{the consequences of acting on incorrect answers} (P6).
Given the museum's vital role in educating and informing visitors, it is crucial to mitigate the risk of inaccurate or `hallucinated' information from LLMs and guarantee the reliability and standardization of the content provided.
A proactive approach is to incorporate user feedback as a cornerstone for the continual refinement of LLMs with accurate and preferred information~\cite{Christiano2017deep}.
Users could signal potentially unreliable content, thus initiating a verification procedure. Moreover, VR environments offer unique opportunities to observe additional human responses, such as facial expressions, and to utilize these as implicit indicators of content trustworthiness.
Future research could explore designing effective and efficient interactive frameworks that utilize human feedback to ensure the veracity and dependability of LLMs within virtual settings.

\noindent
\textbf{Balance between automatic guidance and flexible travel.}
Our work primarily explores the realm of automatic guidance, with the enhancement through LLMs.
\tool is designed to autonomously determine the path, speed, and acceleration of movement within the virtual environment, instead of allowing users flexible travel. 
This design aims to obviate the need for hand controllers, creating an interaction model that more closely resembles real-world navigation, where no extraneous tools are used to moderate movement.
Nevertheless, in our user study, one participant (P5) indicated discomfort with a pre-set speed as it is \q{is faster than my normal walking pace and induces motion sickness}, highlighting the importance of balancing automatic guidance with individual flexible travel experiences. 
To address this, future work could investigate strategies to merge the efficiency of automated navigation with the intuitiveness and adaptability of personalized travel experiences.
One possible method is to dynamically adjust LLM-recommended paths according to user-specified commands and comfort levels in real time, eliminating the discomfort associated with the virtual navigation process.

\noindent
\textbf{Expand generalizability across diverse museum collection types.} 
The initial application of \tool on predominantly planar objects, such as paintings, was chosen due to their widespread presence in art museum collections and the relative ease of their digital representation and interactive engagement. 
However, extending the capabilities of \tool to encompass a broader array of museum content, including 3D objects like sculptures, installations, or archaeological artifacts, introduces distinct challenges and opportunities for LLM-based interactions. 
These arise from the inherent variation in how visitors interact with and engage in these diverse exhibit types.
For instance, 3D objects demand a multifaceted multi-modal interaction framework, accommodating user needs such as circumnavigating the object, observing it from varied perspectives, and potentially engaging with it in a more tactile manner. 
Integrating guidance for 3D objects into \tool necessitates enhancing LLMs' functionality to undertstand spatial contexts and adeptly process spatial interactions and inquiries.
Future developments could focus on investigating a more comprehensive LLM-based interaction framework that accommodates the unique characteristics and interaction requirements of a varied spectrum of museum exhibits.

\subsection{Beyond Virtual Museums}
This work focuses on the specific scenario of virtual reality museums, while our system demonstrates significant generalizability for application in a broader spectrum of real-world functional contexts.
Herein, we discuss the opportunities and challenges when extending our system beyond virtual reality museums.

When participants experience an LLM-guided tour in virtual reality, some of them point out that our tour guidance system exhibited its true utility in \q{complex functional scenarios such as hospitals or airports with clear guidance objectives} (P9).
Our system not only reduces time and human resource costs but also \q{liberates participants from the confines of flat screens} (P7), allowing for a more immersive connection with the physical environment.
Encouragingly, participants suggested that our system could be adapted to cater to specific demographic segments, such as \q{elderly (P9)} and \q{children} (P11).
However, unlike the provision of virtual tour guidance with visual embellishments through VR devices, it is challenging to overlay such extra information in the physical world.
Additionally, virtual reality tends to create compartmentalized experiences and potentially leads to isolation if overlooking the inherently social and collaborative aspects of physical visitor experiences~\cite{eklund2020shoe,hornecker2019human}.
Fortunately, the advances in augmented reality (AR) offer a viable opportunity for seamlessly integrating our visual feedback designs into the tangible environment and bridging the gap between the digital and physical realms with social and collaborative interactions.
We suppose using AR can enhance the guided tour experience by facilitating hands-free interaction with our LLM-enhanced multi-modal feedback.
We are confident in the extensibility of our multi-modal guidance designs to an AR tour scenario, given that our system was designed and implemented within a 3D environment using Unity3D.

However, transferring our designs in real-world scenarios through AR technologies necessitates thoughtful consideration in terms of efficiency of data collection, effectiveness of feedback presentation, privacy safeguards, and adaptability.
First, compared to VR, human behaviors are more complex within the physical environment, which makes it much more difficult to extract human-environment interaction data.
Second, many real-world external factors such as lighting, noise, and human traffic have the potential to disrupt the presentation of feedback and thus influence the user's tour experience.
Third, voice-controlled interactive systems within real-world settings raise social issues such as pertaining to personal privacy and the potential disruption of others' experiences. 
Implicit interaction modalities might be more suitable for real-world scenarios.
Fourth, a comprehensive tour guidance assistance system should accommodate diverse tour environments and continually optimize its performance as users engage with it over time.
In the future, we plan to adapt our current system to AR (through iterative development of context-adaptive feedback visual designs and the establishment of a unified workflow) and evaluate its effectiveness and adaptability with more exploratory tasks in various functional scenarios.

\section{Conclusion}
\label{sec:conclusion}

This work presented \tool, an interactive system designed to provide virtual tour guidance through a combination of multi-modal feedback.
We presented a comprehensive design framework for multi-modal feedback interactions, derived from a formative study.
We designed five distinct multi-modal feedback combinations and utilized large language models to translate user intent into specific guidance-seeking contexts, thereby facilitating the generation of multi-modal feedback responsive to user voice commands.
Through three virtual museum tour examples, we demonstrated the feasibility and versatility of \tool to cater to personalized requirements across various guidance-seeking contexts.
The results of the user study suggest that our system provides rich knowledge, supports natural communications, and enhances user engagement in various exploratory virtual tour experiences, indicating its great potential for expansion into real-world scenarios.

\bibliographystyle{ACM-Reference-Format}
\bibliography{main}

\appendix

\section{Prompt Generation Examples}
\label{sec:appendix}

\begin{longtable}{cl}
    \Xhline{0.7pt}
    \multicolumn{2}{l}{\textbf{User Input: Please help me plan a tour for this museum in 30 minutes.}} \\ 
    \Xhline{0.7pt}
    \multicolumn{1}{c|}{Input (\textcolor{newgreen}\tool)}  & 
    \begin{tabular}[c]{@{}m{0.78\textwidth}@{}}
        \textcolor{newgreen}{Question}: Please help me plan a tour for this museum in 30 minutes.\\
        \textcolor{newgreen}{Position}: (0.0,0.0,0.0)\\
        \textcolor{newgreen}{Landmark}: null\\
        \textcolor{newgreen}{History}: null
    \end{tabular}  \\
    \cline{1-2}
    \multicolumn{1}{c|}{Output (\textcolor{newblue}{\textit{Classifier}})} & \textcolor{newblue}{Task Classfication}: {[}'information enhancement'{]} \\
    \cline{1-2}
    \multicolumn{1}{c|}{Output (\textcolor{newred}{\textit{Explorer}})}   & 
    \begin{tabular}[c]{@{}m{0.78\textwidth}@{}}
        \textcolor{newred}{Response}: Sure! With 30 minutes, I recommend starting with an overview of the museum by visiting "Mona Lisa" by Leonardo da Vinci, "The Scream" by Edvard Munch, and "Impression, Sunrise" by Claude Monet. Then, you can explore different styles and cultures by visiting "The Birth of Venus" by Sandro Botticelli, "The Great Wave" by Katsushika Hokusai, and "Section of Goddess of Luo River" by Zhang Zeduan. Finally, end your tour with "Night in Black and Gold, The falling Rocket" by James Abbott McNeill Whistler and "Composition no.10" by Mondrian. Enjoy your tour!\\ 
        \textcolor{newred}{Context}: Recommendation for the tour:\textbackslash n- Mona Lisa\textbackslash n- The Scream\textbackslash n- Impression, Sunrise\textbackslash n- The Birth of Venus\textbackslash n- The Great Wave\textbackslash n- Section of Goddess of Luo River\textbackslash n- Night in Black and Gold, The falling Rocket\textbackslash n- Composition no.10\\ 
        \textcolor{newred}{Landmark}: null\\ 
        \textcolor{newred}{Tasks}: {[}"information enhancement"{]}\\ 
        \textcolor{newred}{Tours}: {[}'painting 000', 'painting 003', 'painting 005', 'painting 007', 'painting 013', 'painting 008', 'painting 020', 'painting 018'{]}
    \end{tabular} \\
    \Xhline{0.7pt}
    \multicolumn{2}{l}{\textbf{User Input: Summarize the tour and suggest the next painting.}} \\ 
    \Xhline{0.7pt}
    \multicolumn{1}{c|}{Input (\textcolor{newgreen}\tool)}  & 
    \begin{tabular}[c]{@{}m{0.78\textwidth}@{}}
        \textcolor{newgreen}{Question}: Summarize the tour and suggest the next painting.\\
        \textcolor{newgreen}{Position}: (18.0, 2.0, 0.0)\\
        \textcolor{newgreen}{Landmark}: "painting 000"\\
        \textcolor{newgreen}{History}: {[}'painting 005', 'painting 003', 'painting 000'{]}
    \end{tabular}  \\
    \cline{1-2}
    \multicolumn{1}{c|}{Output (\textcolor{newblue}{\textit{Classifier}})} & \textcolor{newblue}{Task Classfication}: {[}'information enhancement'{]} \\
    \cline{1-2}
    \multicolumn{1}{c|}{Output (\textcolor{newred}{\textit{Explorer}})}   & 
    \begin{tabular}[c]{@{}m{0.78\textwidth}@{}}
        \textcolor{newred}{Response}: You have visited some iconic paintings including "Impression, Sunrise" by Claude Monet, "The Scream" by Edvard Munch, and "Mona Lisa" by Leonardo da Vinci. "Mona Lisa" is a masterpiece painted by Leonardo da Vinci during the Renaissance period. It is known for its enigmatic smile and is considered one of the most famous paintings in the world. For your next painting, I suggest visiting "The Last Supper" by Leonardo da Vinci, which is another renowned work by the same artist.\\ 
        \textcolor{newred}{Context}: Recommendation for the tour:\textbackslash n- The Last Supper\\ 
        \textcolor{newred}{Landmark}: Mona Lisa\\ 
        \textcolor{newred}{Tasks}: {[}"information enhancement"{]}\\ 
        \textcolor{newred}{Tours}: {[}'painting 005', 'painting 003', 'painting 001'{]}
    \end{tabular} \\
    \Xhline{0.7pt}
    \multicolumn{2}{l}{\textbf{User Input: Take me to visit the painting named The Birth of Venus.}} \\ 
    \Xhline{0.7pt}
    \multicolumn{1}{c|}{Input (\textcolor{newgreen}\tool)}  & 
    \begin{tabular}[c]{@{}m{0.78\textwidth}@{}}
        \textcolor{newgreen}{Question}: Take me to visit the painting named The Birth of Venus.\\
        \textcolor{newgreen}{Position}: (0.0,0.0,0.0)\\
        \textcolor{newgreen}{Landmark}: null\\
        \textcolor{newgreen}{History}: null
    \end{tabular}  \\
    \cline{1-2}
    \multicolumn{1}{c|}{Output (\textcolor{newblue}{\textit{Classifier}})} & \textcolor{newblue}{Task Classfication}: {[}'navigation'{]} \\
    \cline{1-2}
    \multicolumn{1}{c|}{Output (\textcolor{newred}{\textit{Navigator}})}   & 
    \begin{tabular}[c]{@{}m{0.78\textwidth}@{}}
        \textcolor{newred}{Response}: \{"Introduction": "Certainly! Let's head towards the painting 'The Birth of Venus' by Sandro Botticelli. Follow me as we navigate through the museum to find the painting.", "Tour": ["The Birth of Venus"], "TourID": ["painting 007"]\}\\ 
        \textcolor{newred}{Context}: null\\ 
        \textcolor{newred}{Landmark}: null\\ 
        \textcolor{newred}{Tasks}: ["navigation"]\\ 
        \textcolor{newred}{Tours}: null
    \end{tabular} \\
    \Xhline{0.7pt}
    \multicolumn{2}{l}{\textbf{User Input: Are there any other paintings of the similar style to this painting in this museum?}} \\ 
    \Xhline{0.7pt}
    \multicolumn{1}{c|}{Input (\textcolor{newgreen}\tool)}  & 
    \begin{tabular}[c]{@{}m{0.78\textwidth}@{}}
        \textcolor{newgreen}{Question}: Are there any other paintings of the similar style to this painting in this museum?\
        \textcolor{newgreen}{Position}: (-19.8, 1.4, 17.32)\\
        \textcolor{newgreen}{Landmark}: "painting 007"\\
        \textcolor{newgreen}{History}: ["painting 007"]
    \end{tabular}  \\
    \cline{1-2}
    \multicolumn{1}{c|}{Output (\textcolor{newblue}{\textit{Classifier}})} & \textcolor{newblue}{Task Classfication}: {[}'information enhancement'{]} \\
    \cline{1-2}
    \multicolumn{1}{c|}{Output (\textcolor{newred}{\textit{Explorer}})}   & 
    \begin{tabular}[c]{@{}m{0.78\textwidth}@{}}
        \textcolor{newred}{Response}: "The Birth of Venus" by Sandro Botticelli is an iconic painting from the Early Renaissance period. If you're interested in exploring more artworks with a similar style, I recommend checking out "Mona Lisa" by Leonardo da Vinci and "Last Supper" also by Leonardo da Vinci. These paintings share the Renaissance style and offer a deeper understanding of the artistic techniques and themes of that era.\\ 
        \textcolor{newred}{Context}: Recommendation for the tour:\textbackslash n- Mona Lisa\textbackslash n- The Last Supper\\ 
        \textcolor{newred}{Landmark}: The Birth of Venus\\ 
        \textcolor{newred}{Tasks}: ["information enhancement"]\\ 
        \textcolor{newred}{Tours}: ['painting 000', 'painting 001']
    \end{tabular} \\
    \Xhline{0.7pt}
\end{longtable}

\end{document}